\documentclass{aa}

\usepackage{graphicx}
\usepackage{subcaption}
\usepackage{longtable,tabularx}
\usepackage{comment}
\usepackage{placeins}
\usepackage{orcidlink}
\usepackage{hyperref}
\usepackage{txfonts}
\usepackage[dvipsnames]{xcolor}
\usepackage[normalem]{ulem}
\usepackage{threeparttable}
\begin{document}
\newcommand{\figH}{5cm}

 \newcommand{\rone}{FRB~20121102A\xspace}
 \newcommand{\ronefourseven}{FRB~20240114A\xspace}
 \newcommand{\rthree}{FRB~20180916B\xspace}
 \newcommand{\rsixtyseven}{FRB~20201124A\xspace}
 \newcommand{\roneoneseven}{FRB~20220912A\xspace}
 \newcommand{\delPPA}{$\Delta\mathrm{PPA}$}
 \newcommand{\rmburst}{RM$_{\rm{burst}}$\xspace}
 \newcommand{\rmepoch}{RM$_{\rm{epoch}}$\xspace}
\newcommand{\ppaburst}{PPA$_{\rm{burst}}$\xspace}
\newcommand{\xteeighteen}{XTE J1810$-$197}

   \title{Rapid polarization position angle variations in FRB 20240114A point to a magnetospheric origin}

   \titlerunning{Rapid polarization position angle variations in FRB 20240114A point to a magnetospheric origin}


   \author{Ninisha Manaswini \orcidlink{0009-0005-1319-9586}\thanks{nmanaswini@mpifr-bonn.mpg.de} 
          \inst{1}
          \and
          Dant\'{e} M. Hewitt \orcidlink{0000-0002-5794-2360}
          \inst{2,3}
          \and
          Laura G. Spitler \orcidlink{0000-0002-3775-8291}
          \inst{1}
          \and
          Jason W. T. Hessels \orcidlink{0000-0003-2317-1446}
          \inst{2,3,4,5}
          \and
          Ramesh Karuppusamy \orcidlink{0000-0002-5307-2919}
          \inst{1}
          \and
          Jeff Huang \orcidlink{0000-0002-8043-0048}
          \inst{4,5}
          \and
          Pranav Limaye \orcidlink{0000-0003-3898-5127}
          \inst{1}
          \and
          Lucas Guillemot \orcidlink{0000-0002-9049-8716}
          \inst{6,7}
          \and
          Isma\"{e}l~Cognard \orcidlink{0000-0002-1775-9692}
          \inst{6,7}
}

    \authorrunning{Manaswini, N. et al.}
   \institute{Max-Planck-Institut für Radioastronomie, Auf dem Hügel 69, D-53121, Germany\\
              \email{nmanaswini@mpifr-bonn.mpg.de}
         \and
    {Anton Pannekoek Institute for Astronomy, University of Amsterdam, Science Park 904, 1098 XH, Amsterdam, The Netherlands}
         \and
    {ASTRON, Netherlands Institute for Radio Astronomy, Oude Hoogeveensedijk 4, 7991 PD Dwingeloo, The Netherlands}
        \and
    {Trottier Space Institute, McGill University, 3550 rue University, Montr\'eal, QC H3A~2A7, Canada}
        \and
    {Department of Physics, McGill University, 3600 rue University, Montr\'eal, QC H3A~2T8, Canada}
        \and
    {LPC2E, OSUC, Univ Orl\'eans, CNRS, CNES, Observatoire de Paris, F-45071 Orl\'eans, France}
        \and
    {ORN, Observatoire de Paris, Universit\'e PSL, Univ Orl\'eans, CNRS, 18330 Nan\c{c}ay, France}
   }          

   \date{Received DMY; accepted DMY}

\abstract{
Fast radio bursts (FRBs), thought to originate from magnetars, exhibit diverse polarization properties that provide key constraints on their emission physics and local magneto-ionic environments. Among these, the polarization position angle (PPA) is particularly sensitive to magnetic-field geometry in the emitting region and propagation effects in the magnetosphere and beyond. PPAs in hyper-active repeating FRBs are typically stable across the duration of a burst and, in two well-studied cases, remain stable on timescales of hours to days. In contrast, here we present observations of the repeating source \ronefourseven that show significant burst-to-burst jumps in the PPA.

Using full-Stokes, high–time-resolution observations from the Nan\c{c}ay Radio Telescope (1.1--1.8~GHz) and the Effelsberg 100-m telescope (1.3--1.5~GHz) spanning $\sim$1 year, we measure burst rotation measures (RMs), as well as time-resolved polarization fractions and PPAs at 12 epochs.

The burst RMs remain stable across all epochs, and the emission is predominantly highly linearly polarized, with $\sim81\%$ of bursts showing $L/I > 0.8$, while circular polarization is weaker ($\sim16\%$ with $|V/I| > 0.1$). We find no evidence for Faraday conversion. The PPA exhibits rapid, stochastic variations on timescales from milliseconds to hours, spanning the full range of $\pm90$ degrees during two periods of high activity and $\pm$50 degrees in a third active period. 
We also investigate the statistics of PPA jumps, defined as the difference in PPA between two consecutive bursts. We find that: (1) there is no difference in the distribution of jumps on timescales shorter or longer than 1~s; (2) positive and negative jumps are equally likely; and 3) a jump of $\pm$90 degrees, as expected from, e.g., orthogonal mode jumps, is not more common than any other value. 

No other hyper-active repeaters show this combination of relatively stable RM, consistently high linear polarization, and extreme burst-to-burst PPA variability. The observations disfavor emission from a single fixed region in a rotating magnetosphere and instead point toward either multiple emission regions, or strong magnetospheric propagation effects, and/or propagation effects in the foreground medium, such as plasma lensing.
}


   \maketitle
%

\section{Introduction}

Fast radio bursts (FRBs) are micro- to millisecond-duration, highly energetic, coherent radio transients (see \citealt{petroff2019fast, cordes2019fast} for reviews). Interferometric localization of more than 100 sources\footnote{The FRB Community Newsletter (Volume 06, Issue 12, DOI:10.7298/JSMK-0N23)} has firmly established their extragalactic origin, revealing a diversity of host galaxies and local environments \citep{heintz_host_frb, bhandari_2022_host_frb, gordon_host_frb, bharadwaj_host_frb, law2024deepsynopticarrayscience, sharma_host_frb}. Despite these advances, the nature of their central engine remains debated. Magnetars have emerged as a leading candidate following the detection of FRB-like bursts from the Galactic magnetar SGR~1935+2154 \citep{Bochenek_Galactic_magnetar, chime_galactic_magnetar}, although a wide range of alternative models has also been proposed \citep{Platts_theory_FRBS}.

Only a small fraction of FRB sources are observed to repeat ($\sim2.6\%$; \citealt{2026ApJS..283...34C}). Among the repeater population, the observed burst rate can vary drastically from source to source and over time. Highly active repeating sources can offer particularly valuable insights, since large burst samples enable high-cadence, detailed characterization of time-variable burst properties (both due to intrinsic variability and induced by propagation effects), which in turn characterize the local magneto-ionic environment and the geometry of the emitting region.

Polarimetric studies are particularly powerful for environmental and emission constraints, since FRB emission is often strongly polarized \citep[e.g.,][]{day2020_pol_ASKAP, pandhi2024polarizationproperties128nonrepeating, sherman2024_FRB_pol}. When polarized radiation propagates through magnetized plasma, the plane of linear polarization rotates as a function of wavelength due to Faraday rotation. The magnitude of this effect is characterized by the rotation measure (RM), which is the integrated product of the free-electron density and the line-of-sight magnetic-field component between the source and the observer. Large absolute RM values indicate dense and strongly magnetized environments \citep[e.g.,][]{gregory_gc_magnetar_rm}, while temporal variations in RM trace changes in the intervening magneto-ionic media. After correcting for Faraday rotation, the intrinsic polarization position angle (PPA) describes the orientation of the source's linear polarization vector projected onto the plane of the sky, typically represented modulo $180^\circ$ (i.e., within $\pm90^\circ$). 

Temporal variations in the PPA --- whether within individual bursts (on millisecond timescales), between bursts during an observing session (minutes to hours), or across activity windows separated by months to years --- provide a unique probe of the emission-region geometry and magneto-ionic environment.
In the case of polar cap emission in radio pulsars, the resulting polarization-angle evolution can be interpreted using the rotating vector model (RVM) \citep{1969ApL.....3..225R}, wherein the observed S-shaped PPA swing versus rotational phase reflects the projection of the dipolar magnetic field as the beam sweeps across the observer’s line of sight (LOS).
While many pulsar PPA profiles approximately follow the RVM, deviations can arise from polarization-mode superposition and propagation effects, although highly linearly polarized single-pulse emission can often recover an RVM-like PPA swing \citep[e.g.,][]{Karastergiou_2009,Oswald_2023_partial_coherence,Johnston_et_al_high_LP_RVM}.

Radio-emitting magnetars provide a complementary point of comparison, exhibiting substantial temporal variability in their polarization properties. Observations of several radio-emitting magnetars have revealed PPA behaviour ranging from relatively flat swings to pronounced and evolving variations between epochs, accompanied in some cases by changes in linear and circular polarization \citep{10.1111/j.1365-2966.2007.11622.x,Champion_et_al_J1818_2020,Lower_et_al_J1818_2021,2012MNRAS.422.2489L,Dai_et_al_J1810_2019}. Observations of \xteeighteen\ have also revealed substantial evolution of its polarization properties between epochs and complex PPA behaviour within individual pulses \citep{10.1111/j.1365-2966.2007.11622.x,Lower_2024}.

This provides a useful framework for interpreting FRB PPAs, where the observed phenomenology broadly spans structured intra-burst evolution, burst-to-burst variability, and longer-term population behavior. For the general non-repeating population, large-scale studies with DSA-110 \citep{sherman2024_FRB_pol} and ASKAP \citep{Scott_2025} show that single-burst PPAs are predominantly flat or exhibit sub-microsecond microstructure, but they inherently lack the multi-burst baseline needed to probe temporal, burst-to-burst evolution. Studies of repeating FRBs have revealed both structured PPA evolution within individual bursts and substantial burst-to-burst variability \citep{2020Natur.586..693L, Hilmarsson_2021}, while other repeaters show relatively stable PPAs across bursts and observing epochs \citep{Michilli_2018,bethapudi2025constrainingoriginlongterm}. These diverse behaviors suggest that PPA evolution can probe both the geometry of the emitting region and propagation or polarization-mode effects.

Measurements of the linear polarization fraction (LP) and circular polarization fraction (CP), together with their RM, provide complementary diagnostics for distinguishing between intrinsic and propagation-induced polarization effects. Depolarization---a reduction in the linear polarization fraction at lower frequencies which can arise from RM fluctuations in an inhomogeneous magnetized medium \citep{1966MNRAS.133...67B}, which has been reported in repeating FRBs \citep{2022Sci...375.1266F}. Variations in LP or CP, and correlations between these quantities and RM, can also indicate processes such as Faraday conversion \citep{KumarFC} or polarization-mode mixing \citep{OPM_OG}, thereby helping to distinguish propagation-induced phenomena from intrinsic geometric changes in the emission region.

FRBs exhibit a wide range of time- and frequency-dependent polarization properties, spanning bursts that are nearly $100\%$ linearly polarized to sources with weak or undetectable polarization \citep{pandhi2024polarizationproperties128nonrepeating, sherman2024_FRB_pol}. In addition, RM variability has emerged as a key probe of the dynamic local magneto-ionic environments of repeating FRBs \citep{Xu_2022, Anna_Thomas_RM_reversal, 2025ApJ...982..154N}.  Exploring such polarization studies is essential for distinguishing intrinsic changes in the emission geometry from propagation-induced effects and for refining physical models of the FRB emission mechanism.

In Jan 2024, the Canadian Hydrogen Intensity Mapping Experiment/Fast Radio Burst (CHIME/FRB) collaboration announced the discovery of the repeater \ronefourseven (dispersion measure, DM = 527.7\,pc\,cm$^{-3}$) \citep{2026ApJ...997..334S}, which was subsequently observed intensively by radio telescopes worldwide, leading to the detection of thousands of bursts across a wide range of radio frequencies \citep{2026arXiv260216409U,zhang2026investigatingfrb20240114afast,2024ATel16432....1O,2024ATel16565....1O, Panda_2025,2024ATel16547....1P,2024ATel16597....1H,2024ATel16599....1J,limayer147}. 
MeerKAT first localized the source to arcsecond precision \citep{10.1093/mnras/stae2013}, and the PRECISE collaboration later refined this to milliarcsecond precision using the European VLBI Network (EVN) \citep{bhardwaj2025hyperactivefrbpinpointedsmclike}. These observations associate \ronefourseven with a star-forming dwarf galaxy at a redshift of $z = 0.1300 \pm 0.0002$ \citep{bhardwaj2025hyperactivefrbpinpointedsmclike}, making it the first known FRB located in a satellite galaxy within a larger galactic system --- similar to the Small Magellanic Cloud and Milky Way. High-resolution Very Long Baseline Array (VLBA) observations subsequently revealed a compact, non-thermal persistent radio source (PRS) coincident with the burst location \citep{Bruni_2025}. This association places \ronefourseven among the small group of hyperactive repeating FRBs hosted by dwarf galaxies and accompanied by a luminous PRS. Recent ultra-wideband observations with the Murriyang (Parkes) telescope detected more than $5\times10^3$ bursts from this source over a $\sim16$ month interval and revealed episodes of intense burst activity (“burst storms”) together with complex spectral structure. These phenomena have been interpreted as evidence for propagation effects, potentially arising from plasma lensing in the immediate environment of the source \citep{2026arXiv260216409U}. 

In this work, we present a year-long polarization study of this hyperactive repeater \ronefourseven, using the Nan\c{c}ay and Effelsberg Radio Telescopes.
We examine the temporal evolution of its PPA and its relation to the other polarization properties of the bursts, asking whether the observed polarization diversity is consistent with that seen in other populations of FRBs, pulsars, and magnetars, and whether it can be understood within a stable emission geometry or instead requires changes in the emitting regions or magnetospheric configuration. We further consider whether the PPA variations are intrinsic to the emission or modified by propagation through the magneto-ionic environment, and whether, despite the stochastic burst arrival times, the PPA retains a signature of the rotational behavior of the central engine.
A complementary analysis of the spectro-temporal burst properties will be presented in a companion paper (Huang et al., in prep.).

This paper is organized as follows. In Section~\ref {section2}, we describe the details of observations from both telescopes, followed by burst selection and polarization analyses in Section~\ref{section3}. The results are presented in Section~\ref{section4}, and in Section~\ref{section5}, we discuss possible interpretations, focusing on different possible emission mechanisms. Finally, Section~\ref{section6} summarizes our conclusions.

\section{Observation and Search Strategy}
\label{section2}
\begin{table*}
\caption{Polarization properties of \ronefourseven bursts across multiple observing epochs.} 
\label{tab:frbtable_obs}
\renewcommand{\arraystretch}{1.2}
\small
\begin{tabularx}{\textwidth}{l c X X X X X X X}
\hline \hline
Epoch (MJD) & Day & Duration & Selected Events & \rmepoch & $\mathrm{PPA}_{0,65}$ & $\mathrm{PPA}_{0,84}$ & Telescope \\ 
& & (Hrs) & & rad~m$^{-2}$ & Deg & Deg & \\
\hline
60382.3903356482 & 13 March 2024 & 0.96 & 39 & 362 & 32.20 & 56.35 & Nan\c{c}ay \\
60384.3200115642 & 15 March 2024 & 2.0 & 43 & 363 & 40.85 & 62.01 & Effelsberg \\
60385.3820949074 & 16 March 2024 & 0.78 & 7 & 370 & 24.91 & 46.82 & Nan\c{c}ay \\
60386.3803240741 & 17 March 2024 & 0.93 & 7 & 372 & 52.99 & 66.19 & Nan\c{c}ay \\
60398.2966666534 & 29 March 2024 & 1.5 & 9 & 388 & 27.71 & 30.63 & Effelsberg \\
\hline
60505.0597106482 & 14 July 2024 & 0.93 & 11 & 365 & 23.44 & 28.11 & Nan\c{c}ay \\
60506.0517476852 & 15 July 2024 & 0.96 & 9 & 363 & 14.14 & 17.18 & Nan\c{c}ay \\
60515.0280555556 & 24 July 2024 & 1.04 & 10 & 370 & 16.47 & 22.92 & Nan\c{c}ay \\
\hline
60673.6016782407 & 29 December 2024 & 0.85 & 15 & 349 & 31.27 & 53.52 & Nan\c{c}ay \\
60683.5701041667 & 8 January 2025 & 0.92 & 14 & 340 & 46.86 & 51.42 & Nan\c{c}ay \\
60687.5638657407 & 12 January 2025 & 0.85 & 12 & 345 & 43.24 & 53.15 & Nan\c{c}ay \\
60692.5418402778 & 17 January 2025 & 1.04 & 7 & 348 & 34.88 & 61.27 & Nan\c{c}ay \\

\hline

\hline
\hline
\end{tabularx}

\tablefoot{
The table lists the Modified Julian Date (MJD), observing day, number of events that passed the threshold criteria defined in Section~\ref{bs selection}, mean RM, spread of polarization position angle for all the bursts within each epoch from the
mean ($\mathrm{PPA}_{0,65}, \mathrm{PPA}_{0,84}$), and the telescope used at each epoch. $\mathrm{PPA}_{0,65}$ and $\mathrm{PPA}_{0,84}$ represent the 65th and 84th percentiles of $|\mathrm{PPA}_{0}|$, respectively, computed for each epoch.
}

\end{table*}

\subsection{Nan\c{c}ay Radio Telescope}

Since shortly after its discovery, we have monitored \ronefourseven approximately weekly as part of our FRB monitoring campaign, \'ECLAT (Extragalactic Coherent Lights from Astrophysical Transients; PI: D.~M. Hewitt) on the Nan\c{c}ay Radio Telescope (NRT). At 1.4\,GHz, it has a gain of $G \sim 1.4\,\mathrm{K\,Jy^{-1}}$ and a system temperature of $T_{\rm sys} \sim 35\,\mathrm{K}$, corresponding to a sensitivity comparable to that of a 94-m single-dish telescope. 
We performed observations using the EVN PRECISE localization of \ronefourseven, RA (J2000) = $21^{\mathrm h}27^{\mathrm m}39.9^{\mathrm s}$ and Dec (J2000) = $04^{\circ}19^{\prime}46.1^{\prime\prime}$ \citep{bhardwaj2025hyperactivefrbpinpointedsmclike}, with the low-frequency receiver ($1.1-1.8$\,GHz) of the Foyer Optimis\'e pour le Radio T\'elescope system, at a central frequency of 1.484\,GHz. We acquired full-Stokes data (recorded in a linear polarization basis) with the Nan\c{c}ay Ultimate Pulsar Processing Instrument (NUPPI; \citealt{10.1063/1.3615154}) using 32-bit sampling, a temporal resolution of $16\,\mu$s, and a total bandwidth of 512\,MHz, recorded in eight subbands, each with sixteen 4\,MHz frequency channels.

 In this work, we include observations conducted between February 2024 and February 2025, as described in Section~\ref{bs selection}. Typical observing sessions lasted approximately 1 hour and included a noise diode scan for polarization calibration. The data were searched using the \'ECLAT pipeline, described in detail by \cite{10.1093/mnras/stad2847}, in the DM range of $505 - 560$\,pc\,cm$^{-3}$, down to the native time resolution of 16\,$\mu$s. 
Coherent dedispersion of $527.7$\,pc cm$^{-3}$ was applied within each 4-MHz channel. We then used Heimdall to identify candidates with S/N $>$ 7, which were subsequently vetted using the FETCH machine-learning classifier \citep{Agarwal_2020}.
For detected bursts, we extract short filterbank segments from the original 32-bit raw data at native time and frequency resolution, preserving full polarization information. All subsequent polarimetric analyses were performed on these extracted datasets. Table~\ref{tab:frbtable_obs} summarizes the detections and observing epochs that pass the selection criteria (described in Section~\ref{bs selection}).

\subsection{Effelsberg Radio Telescope}

We also observed \ronefourseven on 15 and 29 March 2024 with the 100-m Effelsberg radio telescope using the P217\,mm 7-beam receiver, which covers the frequency range $1.26-1.51$\,GHz, at the coordinates RA (J2000) = $21^{\mathrm h}27^{\mathrm m}39.83^{\mathrm s}$ and Dec (J2000) = $04^{\circ}19^{\prime}46.02^{\prime\prime}$ \citep{10.1093/mnras/stae2013}. The Effelsberg Direct Digitization (EDD) backend \citep{EDD_paper} recorded full-Stokes data in \texttt{PSRFITS} format with 8-bit sampling, a temporal resolution of $51.2\,\mu$s, and a coherent de-dispersion at DM =$527.7\,\mathrm{pc\,cm^{-3}}$ applied to individual frequency channels of width $0.78125$\,MHz.

We used the \texttt{TransientX} software package \citep{men2024transientxhighperformancesingle}, including its inbuilt RFI zapping, to search the \texttt{PSRFITS} data for burst candidates over a DM range of 500–550\,pc\,cm$^{-3}$ with an S/N threshold of 7 at two temporal resolutions: the native resolution of $51.2\,\mu$s and a version downsampled by a factor of 8. For the native-resolution search, boxcar matched filters ranged from $51.2\,\mu$s to 1\,ms, while for the downsampled data, they ranged from $1-20$\,ms. We did not detect any bursts at the native resolution, whereas we identified 52 events in the downsampled search (see section \ref{bs selection}).

\section{Polarization analysis}
\label{section3}

\subsection{Burst selection}
\label{bs selection}

To ensure reliable statistical characterization of polarization properties, we considered only observing epochs with at least 20 detected events. Within these selected epochs, we further restricted our sample to events with sufficiently high S/N (Heimdall S/N $>$ 15). We define an event as the burst with the highest S/N among detections that occur close together in time; in some cases, a single event may consist of multiple closely spaced (sub-)bursts. To ensure reliable polarization measurements for lower-S/N bursts while preserving time resolution for higher-S/N events, we adopted different temporal downsampling strategies depending on event S/N. We downsample events with S/N $\geq 20$ by a factor of 8 in time, and those with $15 \leq$ S/N $< 20$ by a factor of 16. A total of 131 events in 10 epochs satisfy these criteria (see Table~\ref{tab:frbtable_obs}).

We identified three distinct periods of enhanced burst activity: MJD~60382$-$60398, MJD~60505$-$60515, and MJD~60673$-$60692. Each activity phase spans multiple observing days and includes epochs with elevated daily burst rates. We note that additional observing epochs occurred within these broader activity windows; however, we include only those epochs that meet the polarization-selection criteria described above in the present analysis. A comprehensive overview of all observing epochs, burst counts, and spectro-temporal burst properties will be presented in Huang et al. (in prep.).

For the Effelsberg data, we follow a similar methodology. We restrict the polarization analysis to events with S/N $>$ 10 as measured by \texttt{TransientX}. We extract a $0.8$-s archive file centered on the burst arrival time of each burst using the \texttt{TransientX}-reported MJD from the original 30-minute FITS files, using \texttt{DSPSR} \citep{2011PASA...28....1V}. This processing yields an effective time resolution of $97.6\,\mu$s for events with S/N $> 20$, while events with S/N $< 20$ are downsampled by a factor of two to improve the polarimetric S/N. Table~\ref{tab:frbtable_obs} lists the total number of selected bursts for each MJD; in total, 52 events are selected across the two Effelsberg epochs.

\subsection{Polarization calibration}

For the NRT data, we generate full-Stokes burst archive files with \texttt{DSPSR} \citep{2011PASA...28....1V} and calibrate the polarimetry using \texttt{PSRCHIVE} \citep{2004PASA...21..302H}, following the methods in \citep{Guillemot_2023}. We adopt the full \texttt{Reception} model \citep{2006ApJ...642.1004V, 2013ApJS..204...13V}, which incorporates frequency-dependent gain variations and non-orthogonality of the feeds in addition to standard gain and phase corrections. 

To derive the calibration solutions, we use regular observations of the bright pulsar PSR~J0742$-$2822, conducted once or twice per month with the NRT, during which the receiver feed is rotated by $\sim 180^\circ$ over a 1-hour track to generate calibration files (\texttt{pcm.fits}). We then apply burst calibration using the \texttt{pac} command in \texttt{PSRCHIVE}, combining the pcm file from the closest calibration epoch with 3.33-Hz pulsed noise diode scans acquired during each observing session.

For the Effelsberg data, we excise RFI from the burst archive files using \texttt{clfd} \citep{Morello_2018} and calibrate the polarimetry with the \texttt{pac} routine from \texttt{PSRCHIVE}, using a two-minute noise-diode scan acquired prior to the FRB observations. In this case, we adopt the \texttt{SingleAxis} calibration model, which assumes that only differential gain and phase variations exist between the two polarization channels. This calibration corrects for the instrumental effects of the circular-feed receiver.

We verify the accuracy of the Effelsberg polarization calibration by reconstructing the polarization profile of PSR~B0355$+$54, observed at the beginning of the observing session, and cross-matching it with the corresponding profile from the EPN database \citep{1998MNRAS.301..235G}.

\subsection{Rotation measure estimation}
\label{rm_measurement}

For the NRT data, we measure the RM of each burst, \rmburst, using the Stokes \texttt{QU-fit} method \citep{2020ascl.soft05003P}, applied to the time-averaged frequency spectra of the calibrated Stokes parameters, following the manual calibration procedure mentioned in Appendix~\ref{pol_cal_appendix}. We adopt this method, \texttt{QU-fit}, because it utilizes the full frequency-dependent structure of the Stokes $Q$ and $U$ parameters and allows us to selectively weight well-measured frequency channels, thereby providing a more robust estimate of RM for high-S/N bursts.

The model fits the frequency dependence of the normalized Stokes parameters, $Q/L$ and $U/L$, as:
\begin{subequations}
    \begin{equation}
        \frac{Q}{L} = \cos\!\left( 2 \left[ \frac{c^{2} \, \mathrm{RM}}{\nu^{2}} + \phi \right] \right)
    \end{equation}
    \begin{equation}
        \frac{U}{L} = \sin\!\left( 2 \left[ \frac{c^{2} \, \mathrm{RM}}{\nu^{2}} + \phi \right] \right)
    \end{equation}
\end{subequations}
where $L = \sqrt{Q^2 + U^2}$, $c$ is the speed of light, $\nu$ is the observing frequency, and $\phi$ is the intrinsic polarization angle on the plane of the sky at infinite frequency. We perform the fitting using \texttt{curvefit} \citep{2020SciPy-NMeth} independently for each burst, and derive RM uncertainties from the covariance matrix of the nonlinear least-squares fit. To validate our RM measurements, we independently estimate RM values using \texttt{RM-Synthesis} and the \texttt{rmfit} routine in \texttt{PSRCHIVE}, and find consistent results within uncertainties. All RM measurements presented in this work are in the observer frame (topocentric).

For the Effelsberg data, we determine \rmburst using the \texttt{rmfit} routine from \texttt{PSRCHIVE}, which performs a brute-force search over trial RM values to maximize the linearly polarized intensity. We perform the search over the range $-700$ to $0\ \mathrm{rad\ m^{-2}}$, consistent with the sign convention of the P217-7 beam receiver. We could not perform \texttt{QU-fit} on these bursts due to limited bandwidth.

For each observing epoch, we compute the mean RM, \rmepoch, across all bursts using the fitting uncertainties as weights. We initially use both \rmburst and \rmepoch values to correct the \texttt{PSRCHIVE} calibrated Stokes parameters for Faraday rotation with the \texttt{pam} routine in \texttt{PSRCHIVE}, but they do not produce significant differences in the resulting polarization parameters (see Figure~\ref{fig:ppa_rm_comparison}). We therefore use \rmepoch to correct all bursts within a given observing epoch and derive the polarization parameters, including PPAs, using \texttt{pdv}. In contrast, \citet{Michilli_2018} and \citet{Snelders_2023} found that multiple bursts from \rone could be jointly fitted with a single RM and $\phi$ over hours to a few weeks. We attempted a similar joint fit for the bursts observed on MJD~60382, imposing a common RM and $\phi$ across all bursts. However, no satisfactory joint fit could be obtained, providing further evidence that the burst-to-burst PPA variations are genuine and cannot be explained solely by differences in the adopted RM correction.

\section{Results}
\label{section4} 

\begin{figure}[t]
    \centering
    \includegraphics[width=\linewidth]{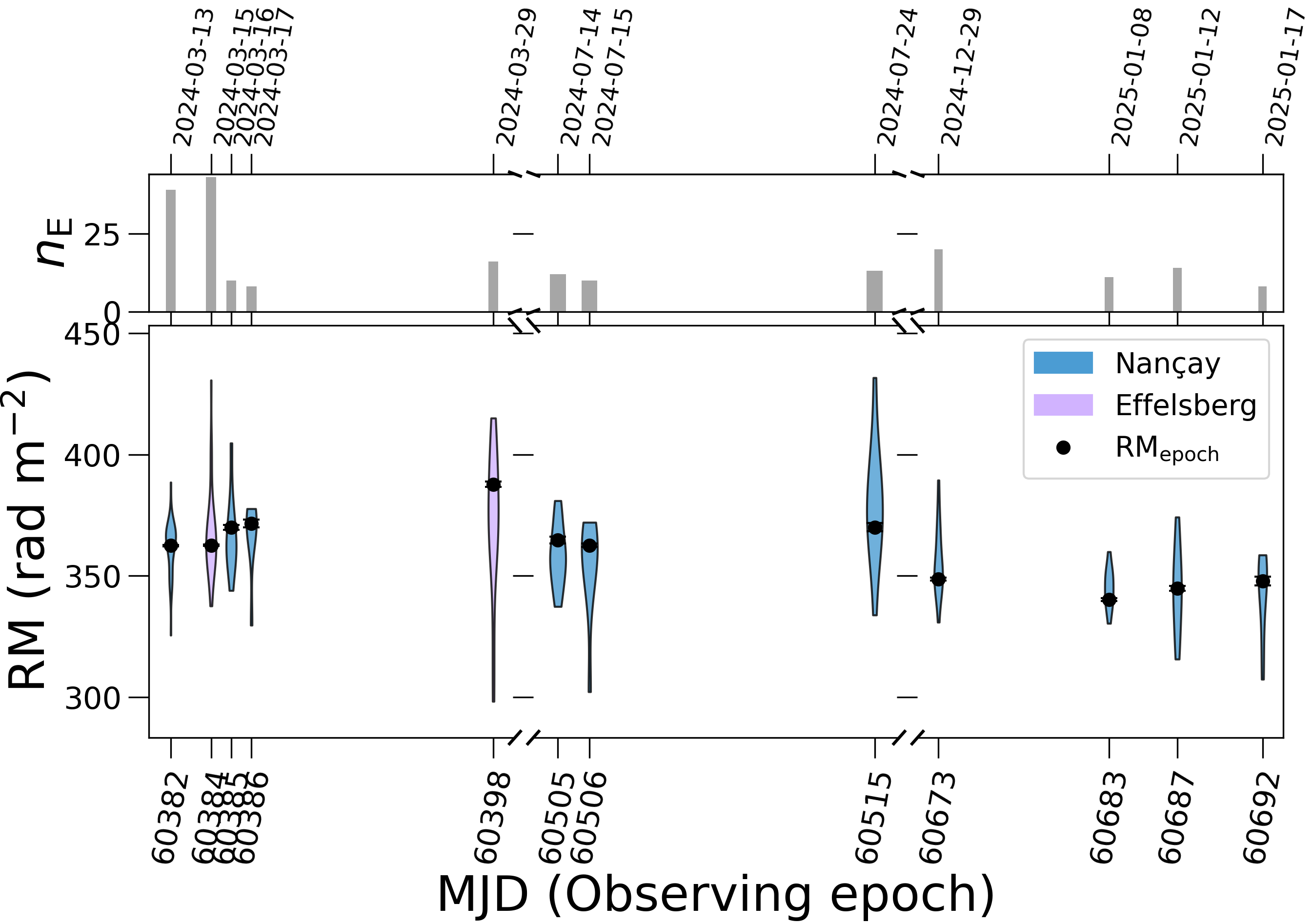}
    \caption{The distribution of \rmburst as a function of observing epoch (bottom). NRT epochs are shown in blue and Effelsberg in pink. For each epoch, a violin plot represents the \rmburst distributions, while black points with error bars indicate \rmepoch and 1-sigma uncertainty. The top panel shows the number of events ($n_{E}$) per epoch used for the RM analysis (i.e., those that satisfy the selection criteria described in Section~\ref{bs selection}). The x-axis is broken to highlight three periods of enhanced burst activity identified by our selection criteria.}
    \label{fig:violin_rm}
\end{figure}

\subsection{Rotation measure}
\label{rm sec2}

The average RM of \ronefourseven remains largely stable across our observational campaign of approximately a year. The \rmepoch values range from $\sim 340 - 387\,$rad\,m$^{-2}$ with modest epoch-to-epoch fluctuations within the standard deviation of $\sim 13\ $rad\,m$^{-2}$. Figure~\ref{fig:violin_rm} shows this RM time series. Our burst selection criteria (Section~\ref{bs selection}) highlighted three periods of enhanced burst activity, separated by a broken time axis.

To investigate possible RM variations on timescales of seconds to minutes, we examined the \rmburst measurements within individual epochs. We show MJD 60384 (Figure~\ref{fig:rm60384}), one of the most densely sampled Effelsberg epochs, as a representative example. The RM measurements at this epoch illustrate the typical behavior observed across the dataset. In general, the RM values within individual epochs show scatter consistent with their uncertainties, with no clear evidence of systematic evolution over time. A small number of outliers are present, typically associated with lower-S/N bursts or limited usable bandwidth for RM determination due to RFI.

As discussed in Section~\ref{rm_measurement}, the use of burst-specific RMs (\rmburst) and epoch-averaged RMs (\rmepoch) yields consistent polarization properties within the measurement uncertainties (Appendix~\ref{ppa_rm_comparison}). We therefore adopt \rmepoch for all subsequent analyses.

\subsection{Polarization fractions}

After correcting for Faraday rotation using \rmepoch (following the same procedure explained in Section~\ref{rm_measurement}), we extracted the Stokes parameters $I$, $Q$, $U$, and $V$, as well as the PPA and its associated uncertainties using \texttt{pdv} from \texttt{PSRCHIVE}. The linear polarization intensity, $L$, was calculated from the quadrature sum of $Q$ and $U$. To account for the positive bias in this measurement, we applied the debiasing method of \citet{2001ApJ...553..341E}:

\begin{equation}
L_{\mathrm{true}} =
\begin{cases}
\sigma_I \sqrt{\left(\dfrac{L_{\mathrm{meas}}}{\sigma_I}\right)^2 - 1},
& \text{if } \dfrac{L_{\mathrm{meas}}}{\sigma_I} \geq 1.57, \\
0,
& \text{otherwise}.
\end{cases}
\end{equation}

We also applied the de-bias correction in calculating $|V|/I$ using \citep{meerkat_pol}:

\begin{equation}
|V|_i =
\begin{cases}
\left|V_{i}\right| - \sigma_I \sqrt{\dfrac{2}{\pi}} & \text{if } \left|V_{i}\right| > \sigma_I \sqrt{\dfrac{2}{\pi}} \\
0 & \text{otherwise}.
\end{cases}
\end{equation}

Using the debiased values, we computed the fractional linear ($L/I$) and absolute circular ($|V/I|$) polarization for each burst. Uncertainties on the polarization fractions were estimated using standard error propagation. In the following, when classifying bursts based on $L/I$ and $|V/I|$, we use the debiased polarization fractions and consider their associated uncertainties.

Across the full sample of 190 bursts, the emission is predominantly highly linearly polarized, with $\sim81\%$ of bursts showing $L/I > 0.8$, while circular polarization is generally weaker, with $\sim16\%$ of bursts exhibiting $|V/I| > 0.1$. 
We further examined the polarization properties across three distinct activity phases. During the first phase, $\sim84\%$ of bursts show $L/I > 0.8$ and $\sim20\%$ show $|V/I| > 0.1$ (109 bursts). In the second phase, these fractions are $\sim70\%$ and $\sim7\%$, respectively (30 bursts). In the third phase , $\sim79\%$ of bursts exhibit $L/I > 0.8$ and $\sim14\%$ show $|V/I| > 0.1$ (51 bursts). While the exact fractions vary between phases, the overall behavior remains consistent: a high degree of linear polarization and circular polarization in a small subset of bursts.

We examine correlations between polarization properties on a burst-by-burst basis. We find no statistically significant correlation between PPA and $L/I$, indicating that variations in the PPA occur largely independently of the degree of linear polarization. In contrast, the circular polarization shows burst-to-burst variability in both magnitude and handedness, occasionally displaying sign reversals within individual bursts.

Taken together, these results show that the emission remains consistently highly linearly polarized. The circular polarization fraction varies between bursts but is not strongly coupled to either $L/I$ or PPA and shows no clear dependence on RM variations. To investigate possible signatures of propagation effects such as Faraday conversion, we examined correlations between $\Delta$RM and $|V/I|$ on a burst-by-burst basis and found no statistically significant correlation across the observing epochs. This behavior disfavors a dominant role for systematic Faraday conversion and instead points toward intrinsic variability in the emission geometry and/or propagation effects in the local source environment. 

\subsection{Polarization position angle}
The PPA was computed in each time sample using the calibrated Stokes parameters, enabling us to examine both intra-burst and inter-burst polarization variations. Comparing absolute PPAs across epochs is challenging due to variations in telescope systematics. No observations of a polarized calibration source with known absolute PPA orientation were available, so an arbitrary offset may exist between the measured PPA values and those that reflect the intrinsic source geometry. We thus focus on relative PPA variations within and across epochs. Specifically, we compute the circular mean of the PPA distribution for each epoch and subtract it from all PPA values observed during that epoch. This procedure preserves the relative PPA variations between and within bursts while removing arbitrary epoch-dependent offsets. 
 
\begin{figure*}[t]
    \centering

    \begin{subfigure}{0.9\linewidth}
        \centering
        \includegraphics[width=\linewidth]{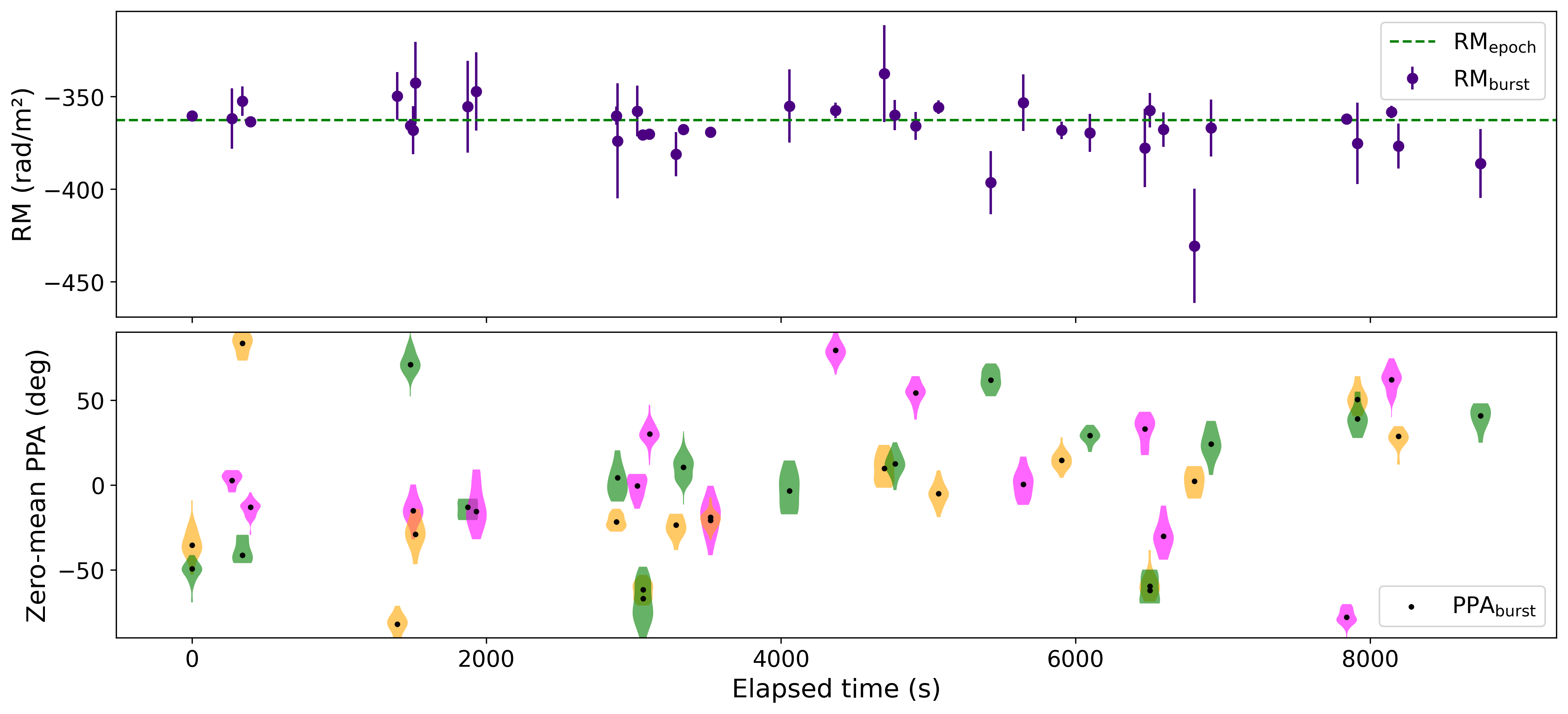}
        \caption{}
        \label{fig:rm60384}
    \end{subfigure}

    \vspace{0.3cm}

    \begin{subfigure}{0.9\linewidth}
        \centering
        \includegraphics[width=\linewidth]{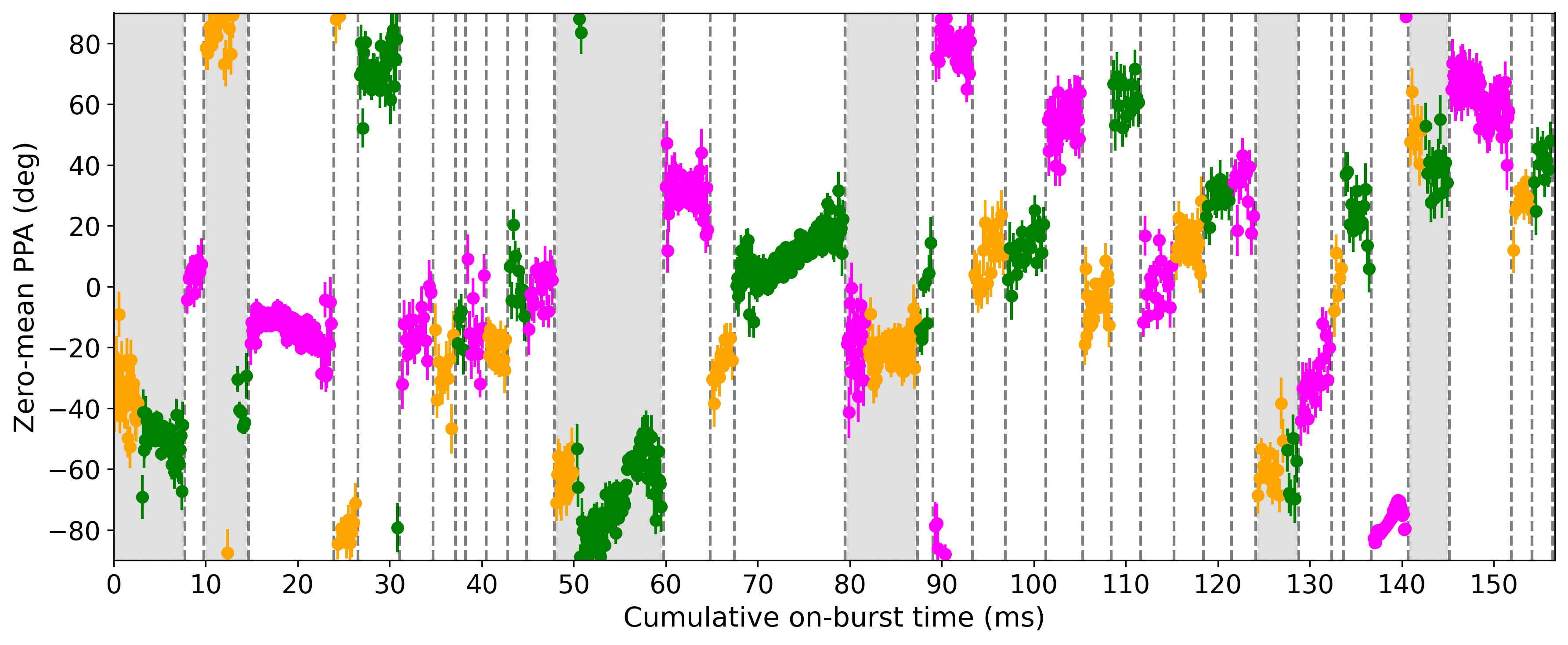}
        \caption{}
        \label{fig:ppa60384}
    \end{subfigure}
\caption{MJD~60384. 
(a) \textit{Top panel:} Burst-specific RMs ($RM_{\mathrm{burst}}$) are shown as violet points as a function of observing time (in seconds since the arrival of the first burst). The weighted-average value for the epoch ($RM_{\mathrm{epoch}}$) is indicated by the green horizontal dashed line. 
        \textit{Bottom panel:} Zero-mean PPAs of individual bursts as a function of observing time. The weighted-average PPA of each burst ($PPA_{\mathrm{burst}}$) is shown as black points, while the spread within each burst is represented by violin plots. Different colors distinguish individual bursts mapped to the bottom plot.
        (b) The PPA of individual bursts are concatenated sequentially in time, preserving their chronological order. The PPAs are shown as a function of cumulative burst duration displayed on a time-sample basis. Different colors distinguish individual bursts, repeated cyclically for clarity. Vertical gray dashed lines mark the boundaries between adjacent bursts, and shaded gray regions indicate cases where consecutive bursts occurred within a 1-second interval.}
\end{figure*}

\begin{figure*}[t]
\centering

\begin{subfigure}{0.48\textwidth}
    \centering
    \includegraphics[width=\linewidth]{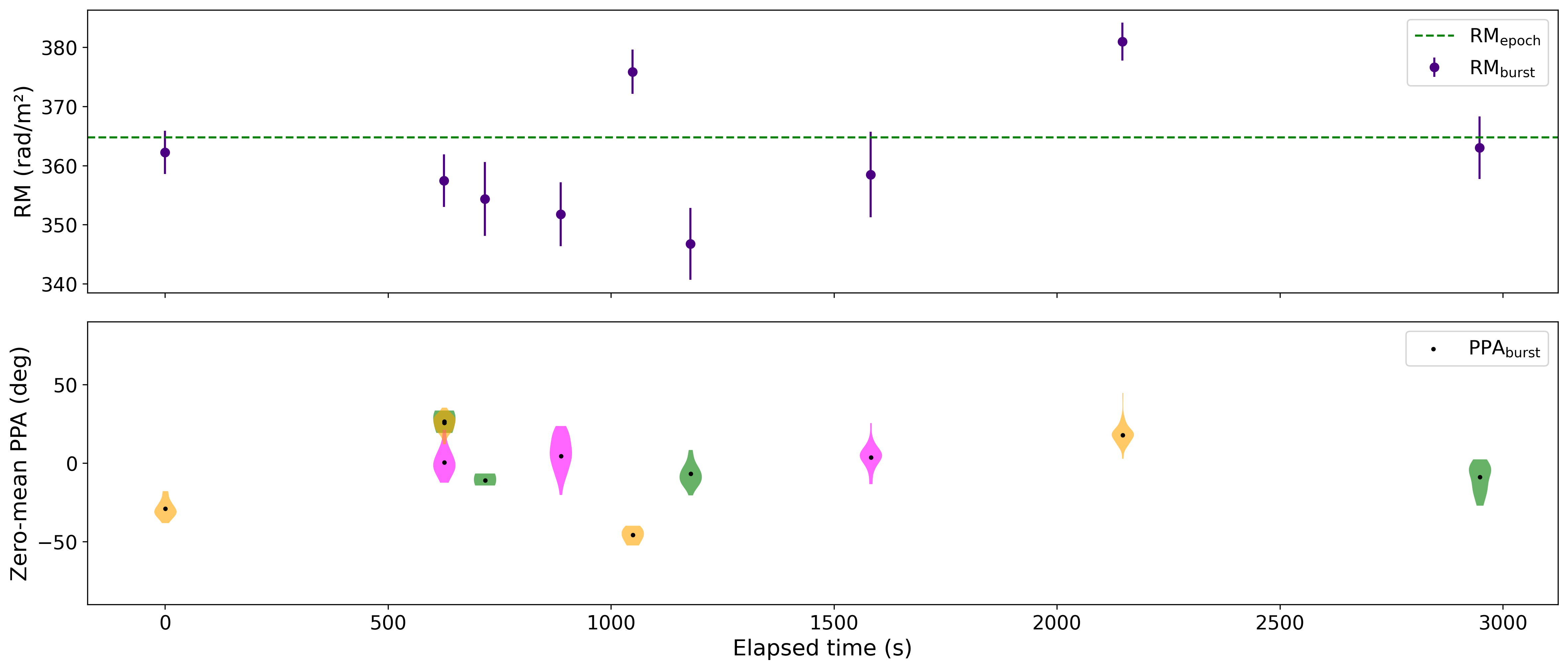}
    \caption{MJD 60505}
    \label{fig:60505_top}
\end{subfigure}
\hfill
\begin{subfigure}{0.48\textwidth}
    \centering
    \includegraphics[width=\linewidth]{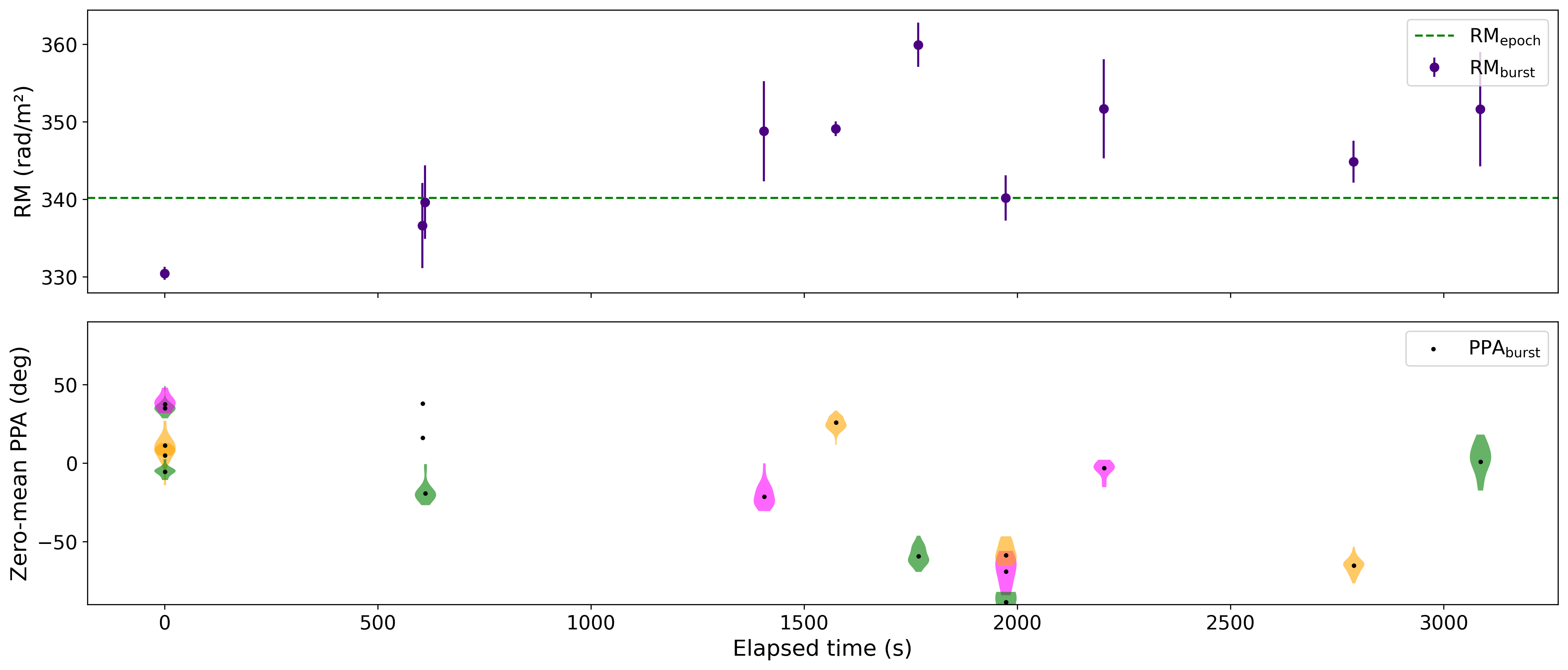}
    \caption{MJD 60683}
    \label{fig:60683_top}
\end{subfigure}

\vspace{0.3cm}

\begin{subfigure}{0.48\textwidth}
    \centering
    \includegraphics[width=\linewidth]{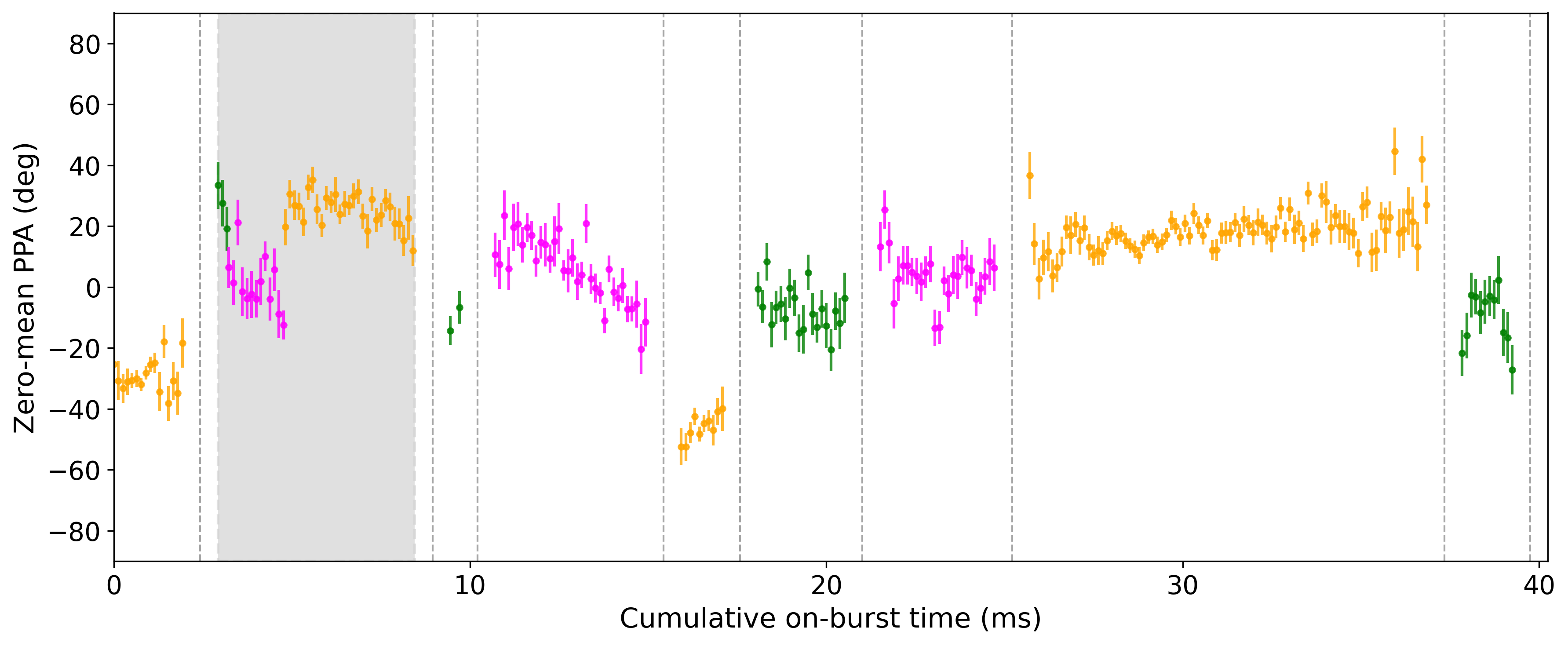}
    \caption{MJD 60505}
    \label{fig:60505_bottom}
\end{subfigure}
\hfill
\begin{subfigure}{0.48\textwidth}
    \centering
    \includegraphics[width=\linewidth]{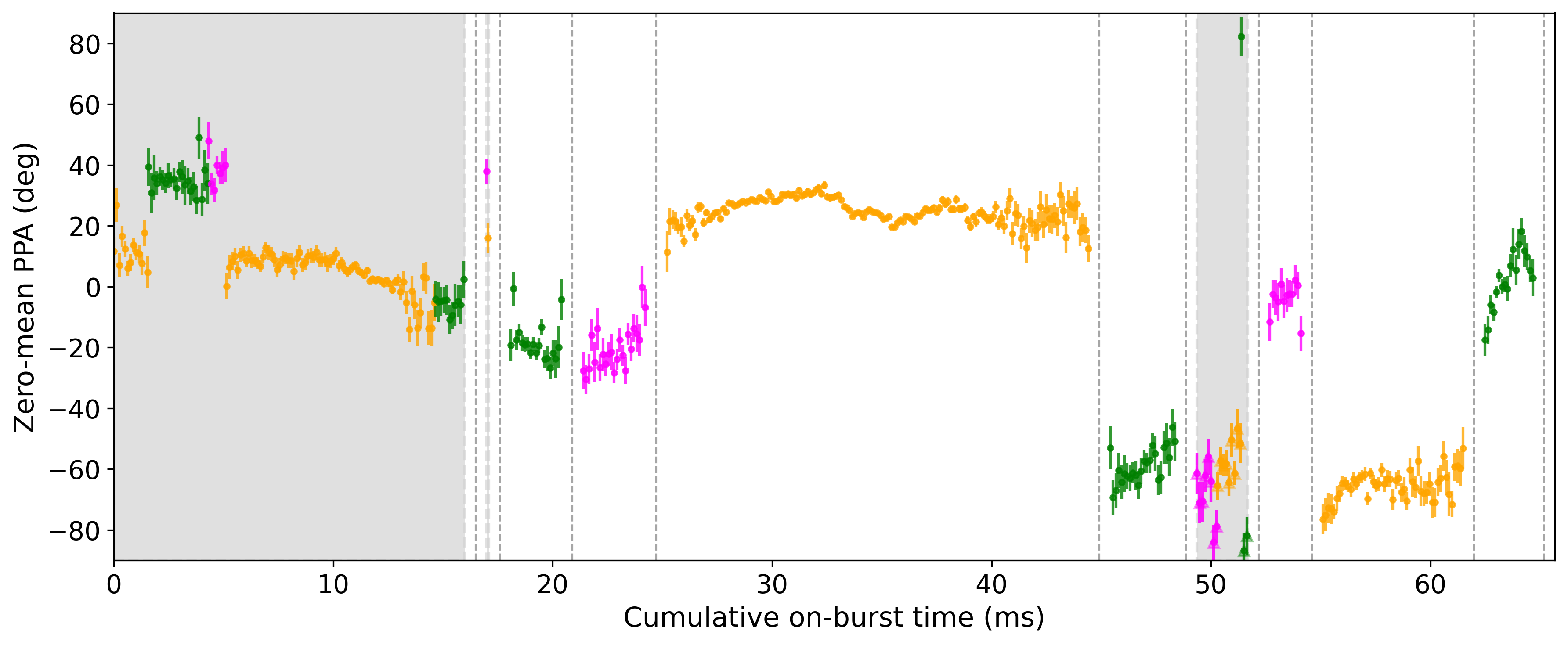}
    \caption{MJD 60683}
    \label{fig:60683_bottom}
\end{subfigure}

\caption{Polarization properties for two representative epochs (MJD~60505 and MJD~60683).
\textbf{(a,b)} Representative epoch from the second phase of high activity.
\textbf{(c,d)} Representative epoch from the third phase of high activity, showing changes in PPA, LP, and CP within individual bursts.}

\label{fig:ppa_compare_epochs}

\end{figure*}

Figures~\ref{fig:ppa60384}, \ref{fig:60505_bottom}, and \ref{fig:60683_bottom} illustrate representative epochs, where we concatenate the time-resolved PPAs of individual bursts in chronological order. Vertical dashed lines mark the boundaries between bursts, while shaded gray regions indicate cases where consecutive bursts occurred within 1~s. The 1~s threshold is motivated by the bimodal waiting-time distribution \citep{junshuo_r147_fast}; 1~s is roughly the minimum between the short and long burst separations. 

\begin{figure*}
\centering

\begin{subfigure}{0.33\textwidth}
\includegraphics[width=\linewidth]{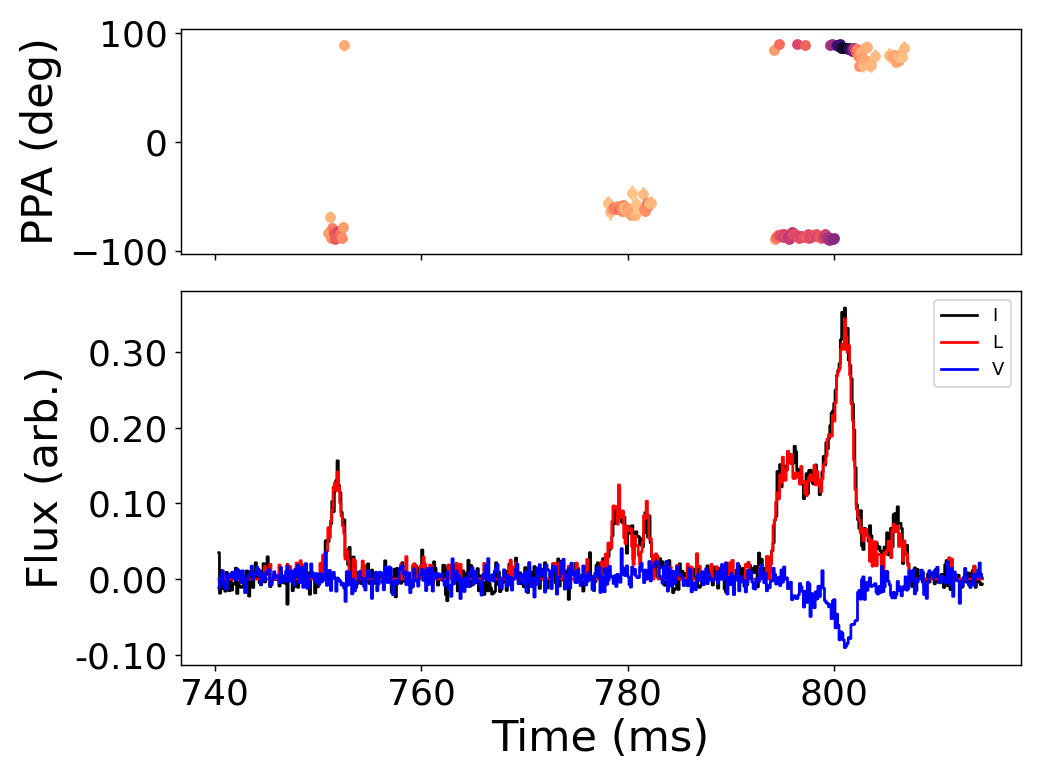}
\caption{}
\end{subfigure}
\begin{subfigure}{0.33\textwidth}
\includegraphics[width=\linewidth]{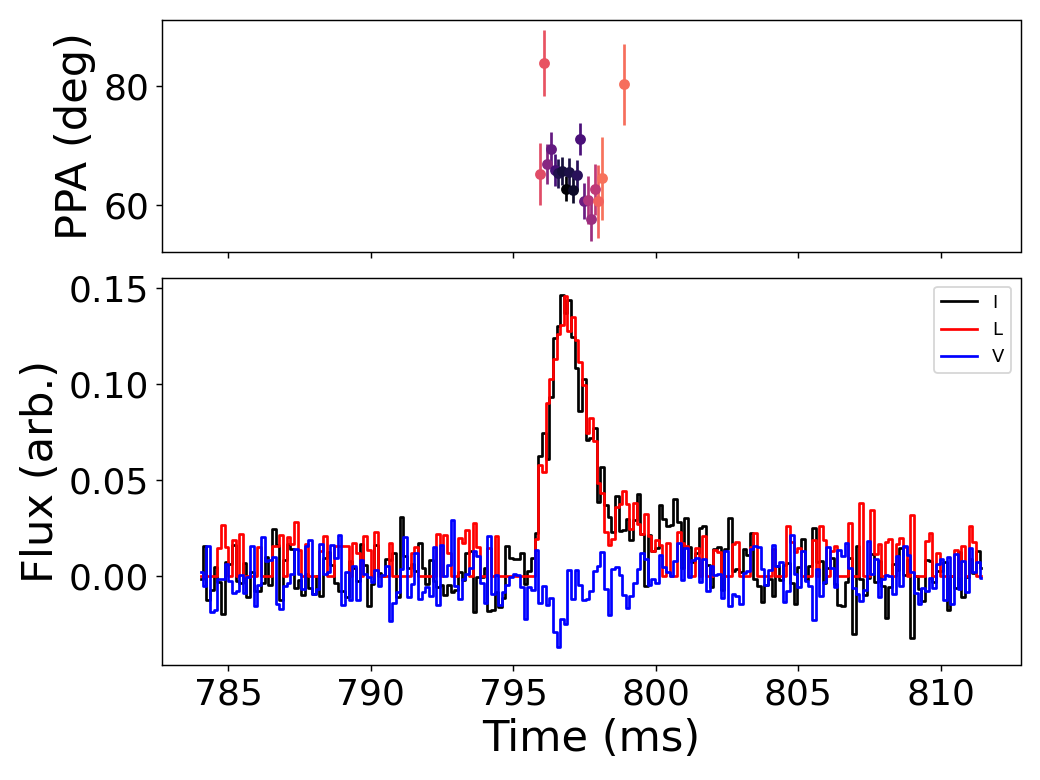}
\caption{}
\end{subfigure}
\begin{subfigure}{0.33\textwidth}
\includegraphics[width=\linewidth]{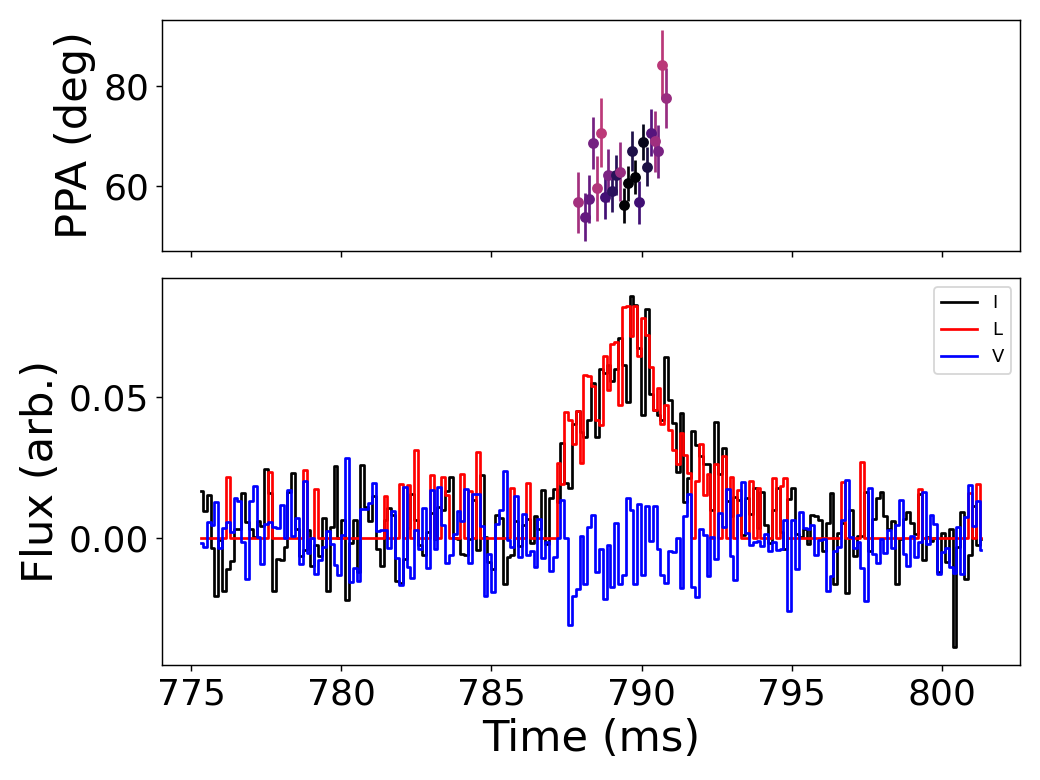}
\caption{}
\end{subfigure}


\begin{subfigure}{0.33\textwidth}
\includegraphics[width=\linewidth]{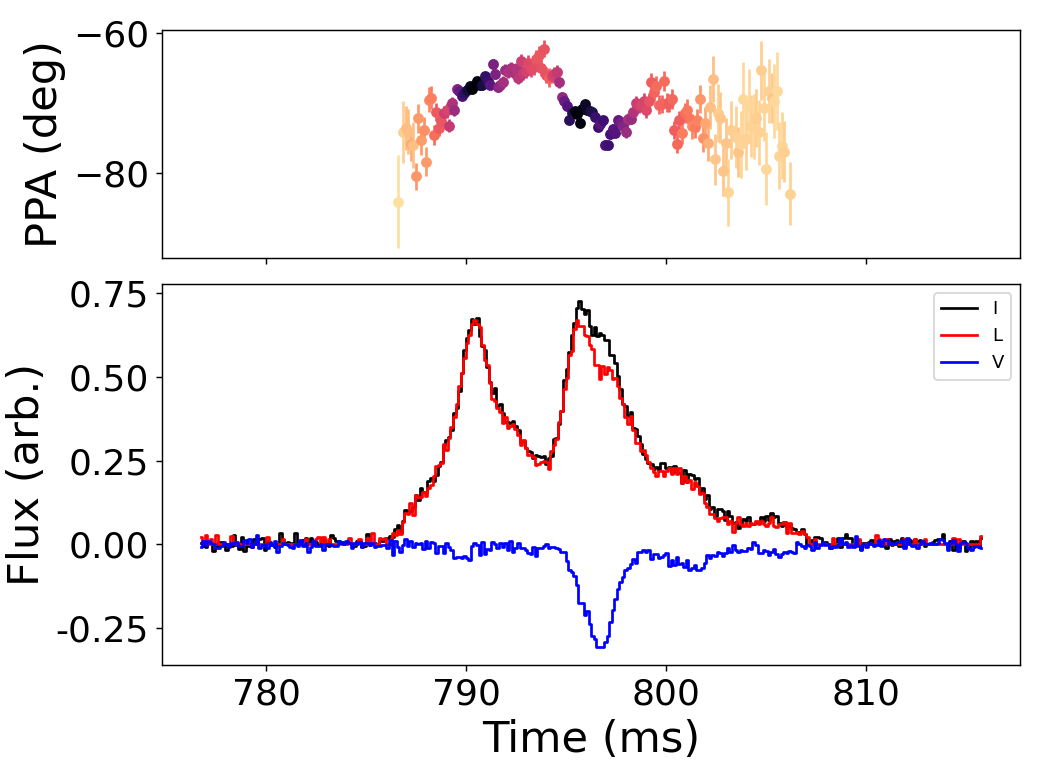}
\caption{}
\end{subfigure}
\begin{subfigure}{0.33\textwidth}
\includegraphics[width=\linewidth]{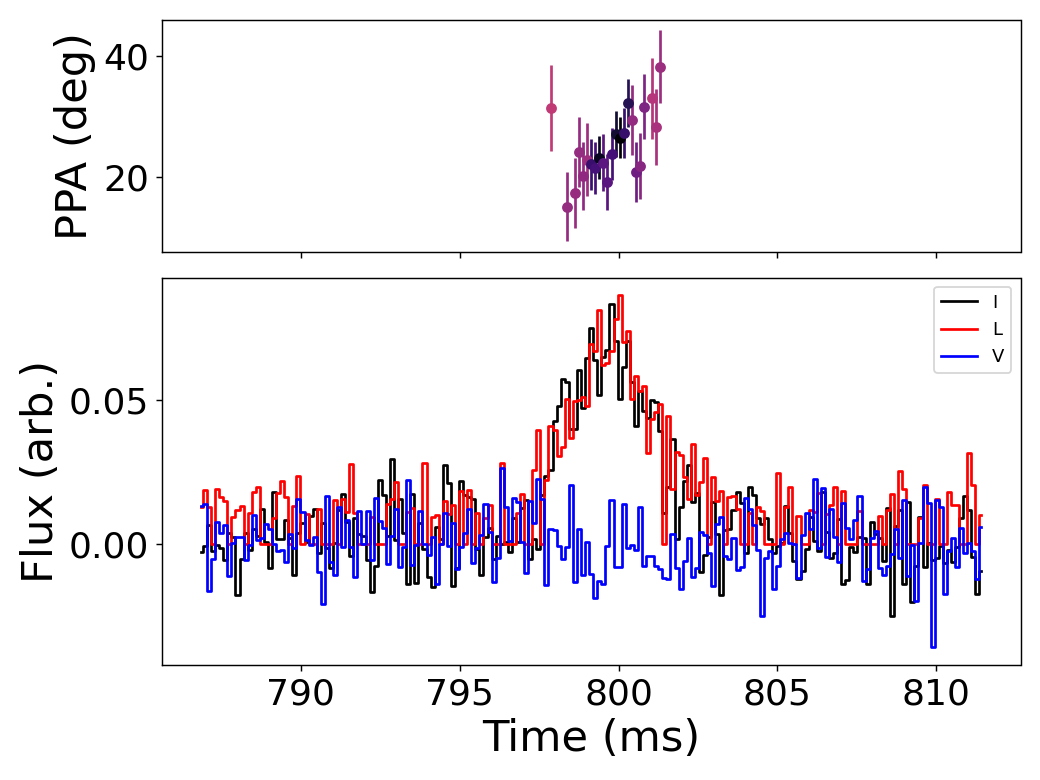}
\caption{}
\end{subfigure}
\begin{subfigure}{0.33\textwidth}
\includegraphics[width=\linewidth]{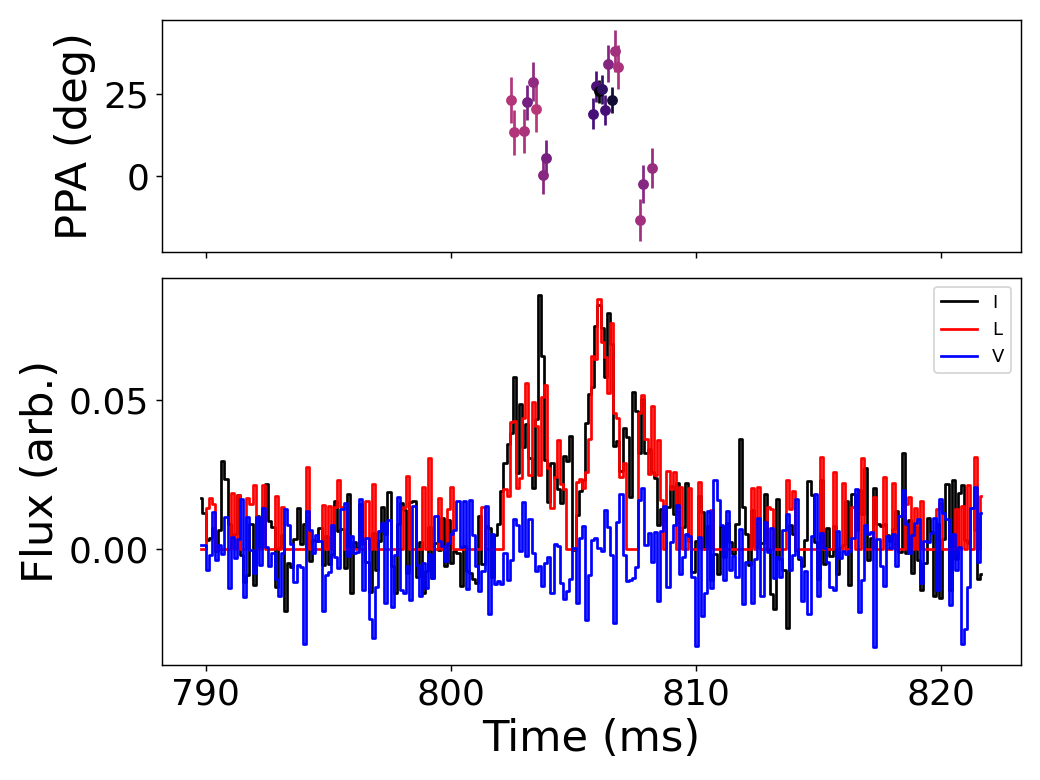}
\caption{}
\end{subfigure}


\begin{subfigure}{0.33\textwidth}
\includegraphics[width=\linewidth]{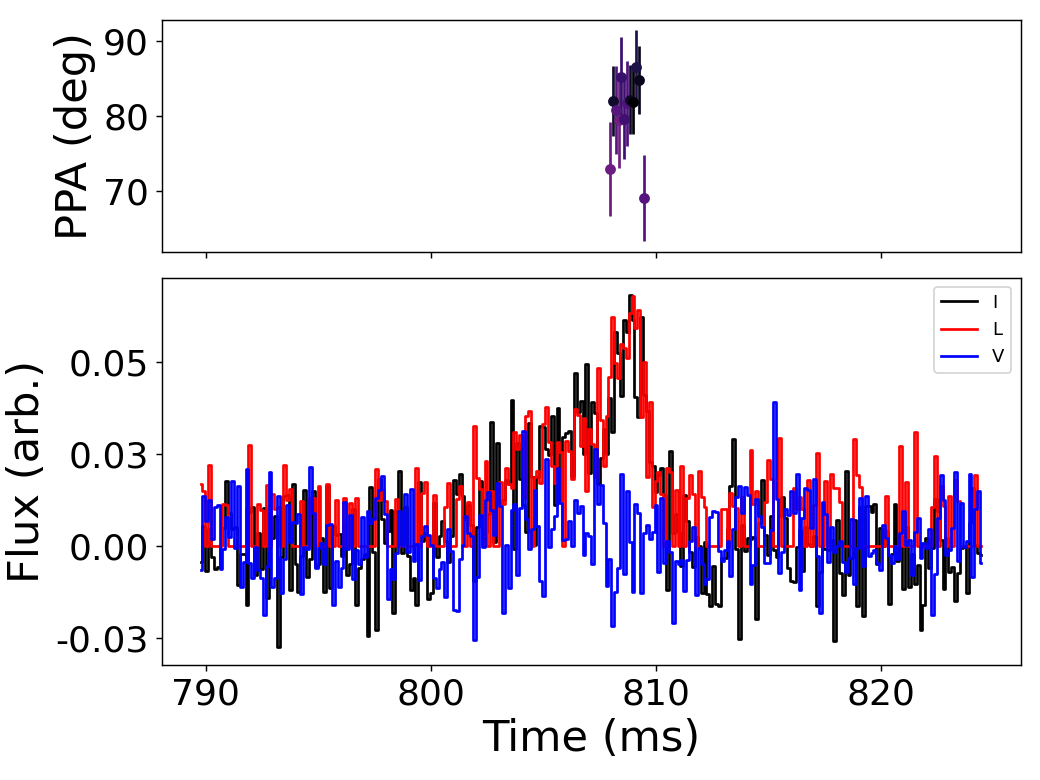}
\caption{}
\end{subfigure}
\begin{subfigure}{0.33\textwidth}
\includegraphics[width=\linewidth]{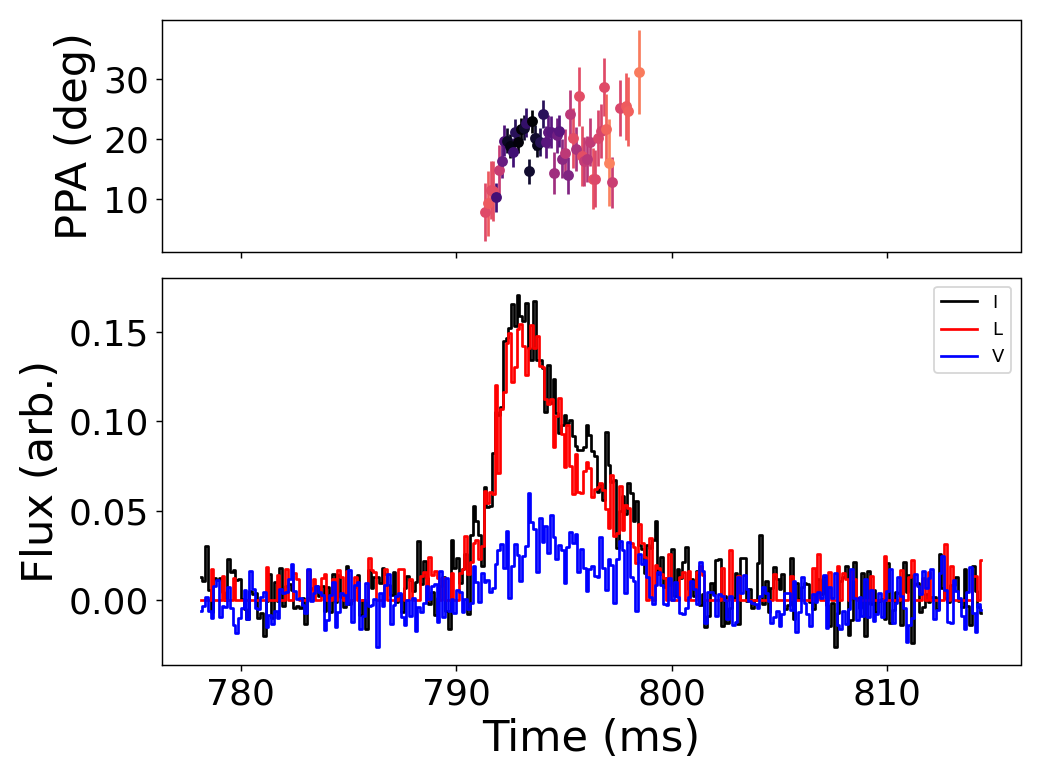}
\caption{}
\end{subfigure}
\begin{subfigure}{0.33\textwidth}
\includegraphics[width=\linewidth]{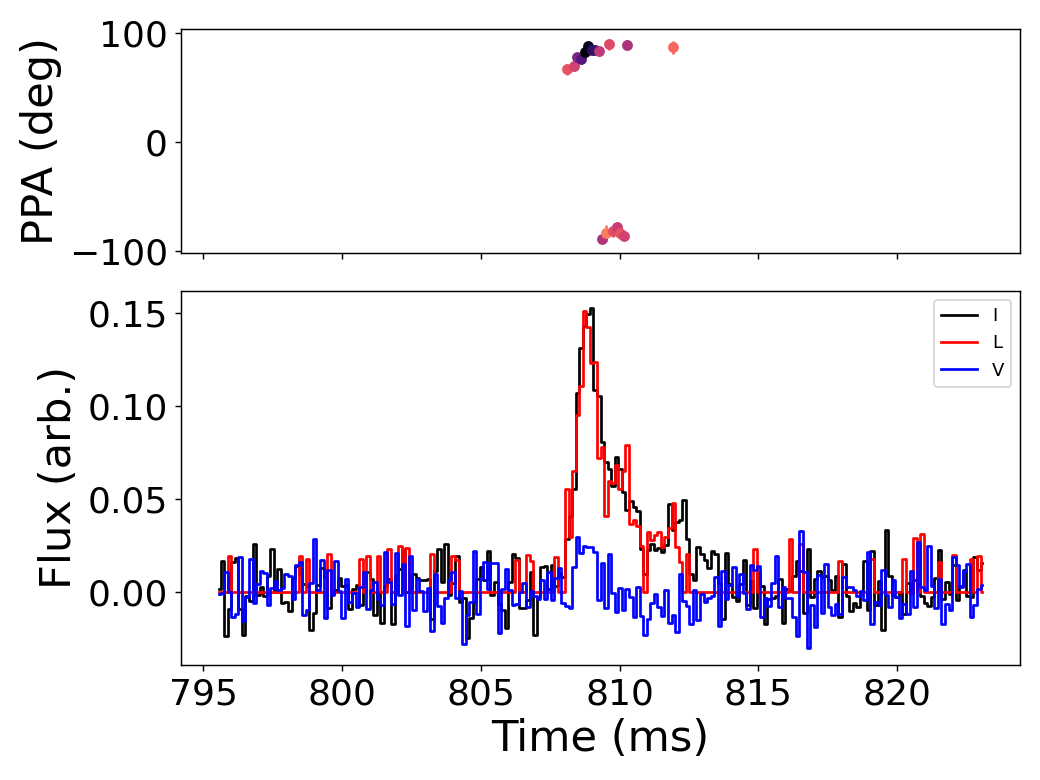}
\caption{}
\end{subfigure}

\caption{Time-resolved polarization properties of nine bursts detected on MJD~60683. For each burst, the top panel shows the PPA as a function of time (without applying any zero-mean referencing or epoch-wise PPA shifting). The PPA measurements are color-coded by the S/N of each time sample. The bottom panel displays the corresponding time series of total intensity (Stokes I, black), linear polarization intensity (L, red), and circular polarization intensity (Stokes V, blue).} 
\label{fig:60683_stacked_ts}
\end{figure*}

Within individual bursts, the PPA is typically flat, although gradients ($\lesssim 40^\circ$) are occasionally observed (Figure~\ref{fig:60683_stacked_ts} - for e.g. see panel d,h, etc.), together with variations in both linear and circular polarization fractions. In rare cases, abrupt $\sim90^\circ$ changes also occur over timescales of a few milliseconds within closely spaced bursts. In MJD~60382, these transitions are particularly evident within consecutive bursts happening within 15~ms (see in the appendix, e.g., the sharp changes in the second gray background around 75 ms in the bottom plot of Figure~\ref{fig:ppa60382}).

The concatenated PPA sequence in Figure~\ref{fig:ppa60384} shows portions where the PPA appears to evolve smoothly across adjacent bursts. If this behavior were a direct consequence of a persistent, ordered magnetic-field configuration, one might expect such smooth evolution to recur systematically across different bursts and observing epochs. However, this behavior is not consistently observed across the dataset and, in MJD 60384, is restricted to particular portions of the plot (around 5700 to 7000 sec in Figure~\ref{fig:rm60384}). We therefore cannot establish from this feature alone that it represents systematic magnetospheric PPA evolution. At the same time, the timescale over which the emission geometry may evolve is unknown, so the $\sim$1000 s duration of the apparent trend does not by itself exclude a geometric origin. Moreover, we find no corresponding systematic evolution in \rmburst over these intervals, and applying \rmburst rather than \rmepoch corrections does not remove the observed PPA differences (Section 3.3). Thus, RM variability is unlikely to account for the apparent continuity.

For each burst, we compute a burst-averaged PPA, \ppaburst, 
defined as the weighted mean of the PPAs across each time sample, where the weights are given by the inverse variance of the PPA uncertainties. In Figures~\ref{fig:rm60384}, \ref{fig:60505_top}, and \ref{fig:60683_top}, these values are shown as black points, while violin plots represent the intra-burst PPA distributions. While individual bursts typically show internally consistent PPA behavior, the \ppaburst\ values reveal pronounced variations between bursts within the same observing epoch. These representative epochs (Figures~\ref{fig:rm60384}, \ref{fig:60505_top}, \ref{fig:60683_top}) also highlight differences across activity phases, with the second phase (MJD~60505 - Figure~\ref{fig:60505_top}) showing a comparatively narrower spread in PPA than the first and third phases.

To quantify the observed variability, we define two complementary metrics. The first characterizes the spread of \ppaburst for all the bursts within each epoch from the mean:
\begin{equation}
\mathrm{PPA}_0 = \mathrm{PPA}_{\rm{burst,}i} - \langle \mathrm{PPA_{burst}} \rangle_{\mathrm{epoch}},
\end{equation}
where $\langle \mathrm{PPA_{burst}} \rangle_{\mathrm{epoch}}$ is the circular average \ppaburst per epoch.
The second measures short-timescale polarization variability between consecutive bursts (that pass our detection and selection criteria),
\begin{equation}
\Delta \mathrm{PPA_{burst}} = \mathrm{PPA}_{\mathrm{burst,}i+1} - \mathrm{PPA}_{\mathrm{burst,}i}.
\end{equation}
All values are wrapped within $\pm90^\circ$ to account for the intrinsic $180^\circ$ ambiguity of polarization angle measurements.

\begin{figure}[t]
    \centering
    \includegraphics[width=\linewidth]{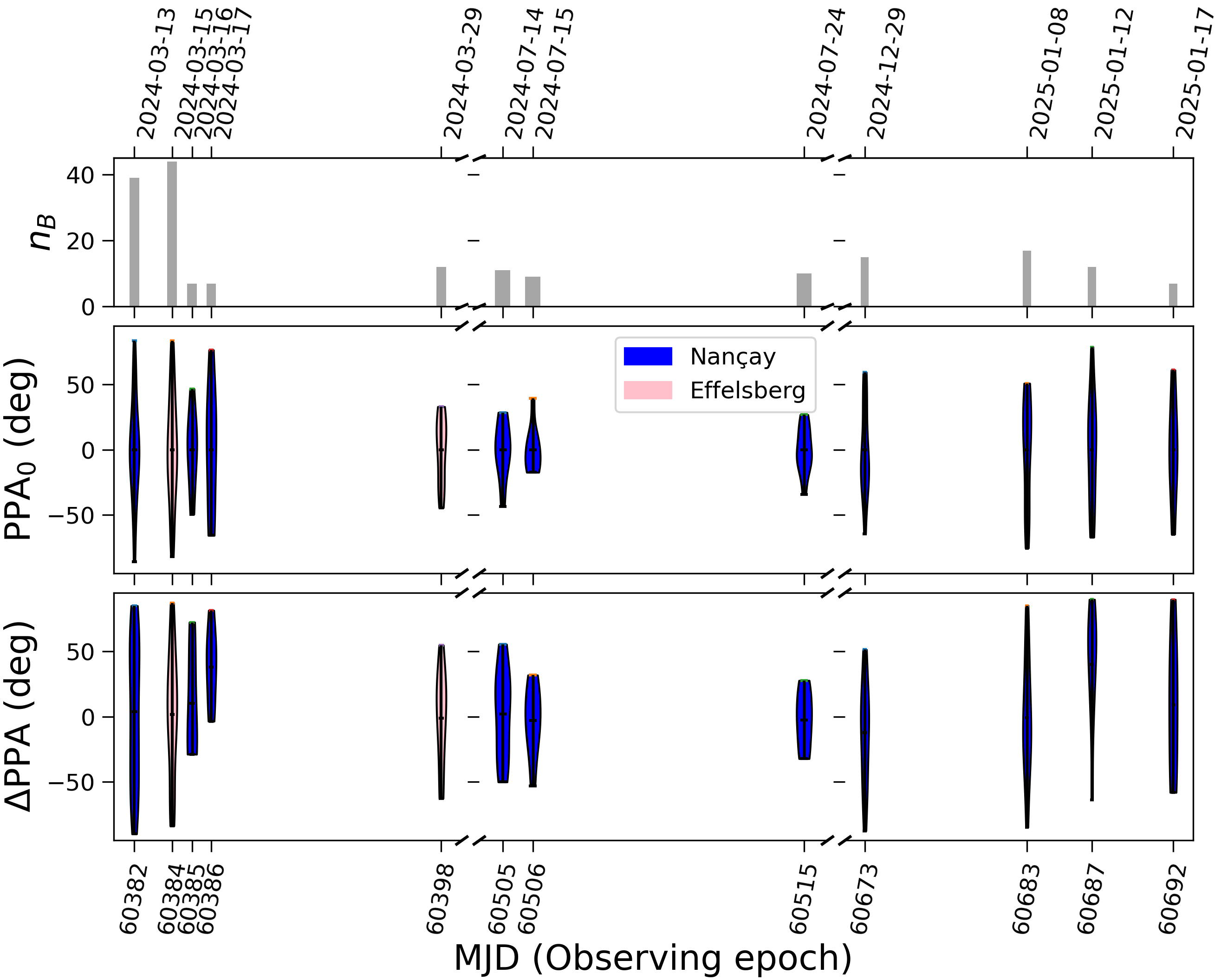}
    \caption{
    Epoch-wise PPA distributions for bursts detected with Nan\c{c}ay and Effelsberg. 
    The top row shows the number of bursts per epoch $\mathrm{n_{B}}$ (gray bars). 
    The middle row presents the zero-mean PPA, $ \mathrm{PPA}_{0}$, obtained by subtracting the epoch mean, and values are wrapped within $\pm 90^\circ$ to account for the $\pm 180^\circ$ ambiguity in polarization angles. 
    The bottom row displays the PPA differences between consecutive time samples, $\Delta \mathrm{PPA}$, computed from the high- and low-resolution datasets; 
    The violin plots show the distribution of $\Delta \mathrm{PPA}$ values, with the mean indicated by a black line. 
    In all violin plots, blue corresponds to Nan\c{c}ay observations and pink to Effelsberg observations.  
    A broken x-axis is used to highlight three phases of higher activity as in Figure \ref{fig:violin_rm}. 
    }
    \label{fig:violin}
\end{figure}

Figure~\ref{fig:violin} shows the distributions of both metrics for all the epochs in Table~\ref{tab:frbtable_obs}. Several key features emerge. 
First, the $\mathrm{PPA}_0$ distributions vary across three high-activity phases, with broader spreads in the first (e.g., MJD~60382; Figure~\ref{fig:ppa60382}) and third (e.g., MJD~60683; Figure~\ref{fig:60683_bottom}) phases and a narrower distribution in the second (e.g., MJD~60505; Figure~\ref{fig:60505_bottom}) phase, indicating reduced variability. 
This difference in $\mathrm{PPA}_0$ behavior mirrors the overall polarization properties discussed earlier, where the second phase also shows a lower fraction of bursts with significant circular polarization. 
Second, the $\Delta\mathrm{PPA}$ distributions are approximately symmetric about zero, indicating no preferred direction for the PPA changes. Similar to the $\mathrm{PPA}_0$ behavior, the $\Delta\mathrm{PPA}$ distribution during the second activity phase is comparatively narrower, indicating reduced burst-to-burst variability relative to the first and third phases.
Third, Large $\sim90^\circ$ jumps, as expected from orthogonal polarization modes, are rare. Fourth, smaller PPA jumps are more common, while larger jumps tend to occur in epochs with higher burst counts. To quantify this trend, we computed the Pearson correlation coefficient between the number of bursts per epoch and the width of the $\Delta\mathrm{PPA}$ distribution, yielding $r=0.67$ ($p=0.017$), indicating a statistically significant, moderate correlation. However, significant variations in the distribution width persist even among epochs in the third phase (MJD~60673, 60683, 60687, 60692) with comparable burst counts (Figure~\ref{fig:violin}), indicating that the observed dispersion is not solely a statistical effect, but reflects real burst-to-burst polarization variability.

To investigate whether burst separation influences polarization angle changes, we performed a two-sample Kolmogorov–Smirnov (KS) test comparing the $\Delta\mathrm{PPA}$ distributions of closely spaced ($\Delta t < 1\,\mathrm{s}$) and more widely separated ($\Delta t \geq 1\,\mathrm{s}$) burst pairs. The test reveals no statistically significant difference between the two distributions, indicating that PPA variability does not depend strongly on the burst separation timescale.

Having established the strong burst-to-burst variations in PPA, we next searched for evidence of an underlying periodic modulation in the epoch with the largest number of detected bursts, MJD 60384. Motivated by the possibility that the PPA may retain information about the rotational phase of the central engine even when burst arrival times are stochastic \citep{kaustabh_ppa_2025}, we applied a Lomb--Scargle periodogram to the time-resolved PPA measurements extracted bin by bin within individual bursts following their approach. We find no significant periodic signal over the range of periods probed by our observations. Thus, despite the large burst-to-burst PPA variations, we find no evidence for a coherent underlying periodic modulation in our dataset.

Overall, the PPA exhibits substantial burst-to-burst variability without a preferred direction or systematic trend, in contrast to sources such as \rone and \rthree, which show stable PPAs.

\section{Discussion}
\label{section5}

The following subsections present a detailed comparison of our results with other known FRB repeaters and discuss the implications in the context of proposed emission and propagation models.

\subsection{Rotation measure}

Recent studies highlight the diversity of Faraday environments among repeating FRBs. \cite{Mckinven_2023a} showed that some repeaters exhibit stable RMs over long timescales, 
while others display secular evolution or episodic changes, with no clear correlation between RM variability and burst activity. 
Similarly, \cite{2025ApJ...982..154N} found that repeaters span a wide range of RM magnitudes, indicating a heterogeneous population shaped by differing local environments.

In this context, \ronefourseven can be contrasted with other well-studied repeaters. 
\rone exhibited an exceptionally large RM that decreased from $\sim10^{5}$ to $\sim10^{4}\,\mathrm{rad\,m^{-2}}$ over several years \citep{Michilli_2018, Hilmarsson_2021}. This strong and evolving RM, together with a DM variation of $\sim25\,\mathrm{pc\,cm^{-3}}$ over 13 years \citep{snelders2025revisitingfrb20121102amilliarcsecond, Li_2025}, has been interpreted as evidence for a dense, highly magnetized, and evolving local environment, such as a young supernova remnant or pulsar wind nebula \citep{Margalit_2018, Li_2025, Hilmarsson_2021}. Some repeaters exhibit more complex behavior, including RM sign reversals \citep{Anna_Thomas_RM_reversal}, implying changes in the line-of-sight magnetic field direction.

Other sources show a more moderate, yet measurable, RM variability. For example, \rsixtyseven exhibits variations of up to $\sim500\,\mathrm{rad\,m^{-2}}$ on $\sim10$-day timescales, 
interspersed with periods of relative stability, which indicate binary interaction \citep{Xu_2022}. 
Several studies also attribute intra-observation RM variations to higher-order propagation effects, 
such as Faraday conversion \citep{KumarFC, Xu_2022}, although these effects may be limited to specific bursts or sub-components.

In contrast, sources such as \rthree show comparatively small absolute RM variations, despite exhibiting noticeable secular trends over time \citep{Mckinven_2023, 2024MNRAS.527.9872G, Bethapudi_2025, 2025ApJ...982..154N}. 
Fluctuations in a comparatively weak, turbulent Faraday-rotating medium can explain this behavior. We also note that \roneoneseven shows near-zero RM \citep{10.1093/mnras/stad2847}.

Over the one-year baseline of our observations, \ronefourseven shows a largely stable RM. The burst-to-burst scatter within individual epochs is consistent with measurement uncertainties, and the observed epoch-to-epoch variations remain within $\sim13\,$rad\,m$^{-2}$ ($\sim3.6\%$ of the mean RM). We therefore find no statistically significant evidence of intrinsic RM variability within or across epochs, with a small number of outliers likely attributable to low-S/N measurements. However, we note that epochs after MJD $\sim60654$ show a gradual decrease in RM at a rate of $\sim 0.9\,\mathrm{rad\,m^{-2}\,day^{-1}}$ \citep{2026arXiv260216409U}. Overall, these results are consistent with previous polarization studies of this source \citep{xie2024polarizationcharacteristicshyperactivefrb, 2026arXiv260216409U}. This relative RM stability suggests a less dynamically evolving Faraday environment compared to sources such as \rone and \rsixtyseven, despite the high burst rate of \ronefourseven. 
In this respect, its temporal RM stability --- though not its absolute RM value --- is more reminiscent of the hyperactive repeater \roneoneseven 
\citep{10.1093/mnras/stad2847, 2023ApJ...955..142Z}.

\subsection{Polarization fractions}

High linear polarization fractions are commonly observed in repeating FRBs, while circular polarization is generally weaker and detected less frequently. Recent population-level studies analyzing a comprehensive sample of 41 repeating FRBs, combining \citep{Li_2025} and \citep{pandhi2024polarizationproperties128nonrepeating}, demonstrate that repeaters exhibit extreme diversity in their linear polarization fractions, ranging from unpolarized upper limits to nearly $100\%$. Rather than converging on a characteristic degree of polarization, this remarkably broad population-level distribution indicates that repeating sources exhibit highly diverse, intrinsically variable emission properties.

Bursts from \rone are typically highly linearly polarized at frequencies above $\sim4$\, GHz \citep{Michilli_2018}, with only a minority exhibiting significant circular polarization \citep{Snelders_2023} and very few showing negligible polarization fractions \citep{Li_2025}. At lower frequencies ($1$--$2$\, GHz), significant depolarization has been reported, which can be naturally explained by spatial depolarization arising from slight density or magnetic variations within the foreground magneto-ionic screen \citep{Plavin_R1_depol}. Similarly, most bursts from \rthree are strongly linearly polarized, with stable polarization properties over hour-long timescales \citep[e.g.][]{Sand_gbt_R3,bethapudi2025constrainingoriginlongterm, 2025ApJ...982..154N}.

In contrast, \rsixtyseven displays more diverse polarization behavior: both linear and circular polarization fractions vary substantially, sometimes on intra-day timescales \citep{Hilmarsson_2021, Xu_2022, Jinchen_R67_fast_pol}. Observations of the active repeater \roneoneseven reveal that while most of its bursts are nearly $100\%$ linearly polarized, approximately $28\%$ in a sample from FAST exhibit a significant circular polarization fraction exceeding $10\%$. Furthermore, this circular component can be extreme, with maximum measured $|V/I|$ reaching up to $\sim\ 70\%$ \citep{2023ApJ...955..142Z}.

The polarization properties of \ronefourseven at $\sim$1.4\,GHz are more similar to those of \rsixtyseven and \roneoneseven than to those of \rone and \rthree. The vast majority of bursts are strongly linearly polarized, with $81\%$ of the bursts showing $L/I > 80\%$, consistent with the behavior commonly observed in repeating FRBs. However, we also observe episodes of enhanced circular polarization within individual bursts (e.g., Figure~\ref{fig:60683_bottom}). The coexistence of high linear polarization with intermittent circular components suggests either intrinsic variability in the emission mechanism or propagation effects acting on short timescales in the local environment.

\subsection{Polarization position angle}

Here we place the PPA behavior of \ronefourseven in the context of other well-studied repeating FRBs. A summary of the PPA variability reported for these sources, including \ronefourseven, is provided in Table~\ref{tab:frbtable}.

Observations of FRB~20180301 revealed the first examples of diverse PPA swings occurring across individual burst envelopes, marking a stark contrast to the flat PPA profiles that were previously thought to characterize repeating FRBs \citep{2020Natur.586..693L}. The smooth $\sim130^{\circ}$ RVM-like swing observed in FRB~20221022A provides another example in which PPA evolution may trace an ordered magnetospheric geometry \citep{McKinven-swing_2024}. Observations of FRB~20220912A show highly variable PPAs frequently accompanying extreme circular polarization fractions of up to $70\%$, which were interpreted by \citet{2023ApJ...955..142Z} as disfavouring relativistic shock-wave models and instead supporting an intrinsic magnetospheric origin, such as coherent curvature radiation. \rsixtyseven also exhibits complex PPA behavior, including variations from burst to burst within a single observing day \citep{Hilmarsson_2021} and occasional $\sim90^{\circ}$ jumps within a single burst \citep{2024ApJ...972L..20N}. These jumps have been interpreted as orthogonal polarization mode transitions, similar to those observed in radio pulsars \citep{1990ApJ...352..258R}. 

On the other hand, several well-studied repeaters exhibit remarkably stable PPAs across bursts and observing campaigns. For example, \rone shows consistent PPAs across multiple observations, despite significant RM evolution \citep{Michilli_2018, Snelders_2023}.
Likewise, \rthree exhibits stable PPAs on timescales of several hours but jumps in PPAs on timescales of days to years \citep{bethapudi2025constrainingoriginlongterm}. Such stability has been interpreted as evidence for a stable orientation of the emitting region with respect to the observer.

\ronefourseven differs from \rone and \rthree (stable PPAs) by exhibiting strong burst-to-burst PPA variability within single epochs, similar to the behavior seen in \rsixtyseven. In a small number of cases, we also observe continuous variation in PPA across the burst envelope, primarily driven by changes in circular polarization (e.g., Figures~\ref{fig:60683_bottom} and \ref{fig:60683_stacked_ts}). In other cases, the PPA exhibits abrupt transitions, or jumps, between discrete values on short timescales. Previous studies have often emphasized $\sim90^{\circ}$ PPA jumps as evidence for orthogonal polarization modes. However, in our sample, the observed PPA changes are not confined to exactly $90^{\circ}$ but span a broader, more continuous range of angles. This behavior is consistent with recent Parkes ultra-wideband observations of \ronefourseven, which also report complex and highly variable PPA structures \citep{2026arXiv260216409U}. High time-resolution FAST studies of this FRB also reveal similar burst-to-burst PPA variability, with the PPA distribution spanning $-90^{\circ}$ to $90^{\circ}$ over timescales exceeding one year \citep{wang2026fastpolarizationcatalogfrb}. 
Given that our analysis preferentially samples epochs with high burst rates, it is plausible that we are probing similar high-activity phases, suggesting a connection between burst activity and enhanced PPA variability (Figure~\ref{fig:violin}). The variations in the $\mathrm{PPA}_0$ distribution across these high-activity phases indicate that the polarization variability is not uniform, but instead reflects changes in the underlying emission or propagation conditions.

The abrupt PPA changes drawn from a wide distribution of values, and not just discrete modes expected from natural plasma modes (see Section~\ref{Faraway model}), observed in \ronefourseven, as well as the stable PPAs observed from \rone and \rthree, place strong requirements on an emission model. Importantly, these intra-burst PPA variations occur despite the absence of measurable long-term RM evolution. This indicates that the observed PPA variability is either caused by small-scale structures within the Faraday-rotating medium 
or intrinsic variability in the emission region and/or propagation effects within the magnetosphere.

Taken together, these comparisons indicate that \ronefourseven combines a relatively stable Faraday environment with rich and rapidly varying PPA behavior at the burst level within epochs over a span of $\sim\ 1$ year. This result raises the question of whether a single underlying population can describe all hyperactive repeaters. 

\subsection{Plasma Lensing}

Propagation through small-scale plasma structures near the FRB source can significantly modify the observed burst properties. In particular, plasma lensing --- refraction caused by localized electron-density inhomogeneities --- can magnify or suppress bursts and introduce complex temporal and spectral structure \citep[e.g.,][]{Cordes_plasmalensing}. If the lensing plasma is magnetized, different propagation paths may accumulate slightly different Faraday rotations, potentially modifying the observed polarization properties. For example, the resulting differential Faraday rotation can lead to overlapping polarization states with nearly orthogonal polarization angles \citep{li2026roleplasmalensingfast}. Such effects can produce rapid PPA swings or apparent $\sim90^{\circ}$ jumps without requiring intrinsically orthogonal emission modes.

Recent observations of \ronefourseven support the presence of such propagation effects. Using ultra-wideband observations with the Murriyang (Parkes) telescope, \citet{2026arXiv260216409U} reported more than $5\times10^{3}$ bursts detected over a $\sim16$-month interval and identified episodes of intense burst activity (“burst storms”) together with highly structured spectral behavior. These phenomena, including recurring narrowband bursts with nearly identical spectral shapes, were interpreted as evidence of extreme scattering events produced by plasma lenses near the FRB source.

The PPA behavior observed in \ronefourseven --- including both gradual swings within bursts and occasional abrupt angle changes --- is qualitatively consistent with the behavior reported by \citet{2026arXiv260216409U} (see also their Extended Figure~12), where similar variations are interpreted in the context of plasma lensing. While our dataset does not allow for an independent verification of plasma lensing, the observed polarization variability is broadly consistent with such a scenario. 

We also note that our selection criteria favor epochs with higher burst rates, corresponding to periods of enhanced activity (``burst storms'') identified by \citet{2026arXiv260216409U}. A direct test would involve investigating the PPA distribution as a function of frequency, since propagation effects are smaller at higher frequencies than at lower ones.

\subsection{Underlying emission mechanism}

FRB emission models fall broadly into two categories: emission occurring well outside the magnetosphere of a compact object (far-away models) and emission originating within the magnetosphere (nearby or magnetospheric models). Below, we compare these models and discuss which scenario best explains the rapid variations in PPA observed in \ronefourseven.

\subsubsection{Far-away models}
\label{Faraway model}
In external-shock scenarios, FRBs are produced when magnetized ejecta launched during a magnetar flare, or by a jet, drive a relativistic forward shock into the surrounding medium, generating coherent synchrotron-maser emission \citep[e.g.][]{2019MNRAS.485.4091M}. In such models, the polarization properties of the radio emission are primarily governed by the structure, orientation, and magnetization of the upstream magnetic field encountered by the shock. The emitted radiation propagates in the natural plasma modes: the extraordinary (X) mode, in which the electric field is perpendicular to both the propagation direction and the upstream magnetic field, and the ordinary (O) mode, in which the electric field lies in the plane defined by the propagation direction and the magnetic field \citep{Plotnikov_2019}.

Particle-in-cell simulations show that for highly magnetized upstream media ($\sigma_q > 1$), the emission is expected to be nearly $100\%$ linearly polarized and dominated by the X-mode \citep{2019MNRAS.485.4091M}. For ordered and approximately axisymmetric magnetic-field geometries, the resulting PPAs are therefore expected to remain stable across bursts on short timescales (minutes to hours), consistent with sources such as FRB~20121102A and FRB~20180916B \citep{Michilli_2018, bethapudi2025constrainingoriginlongterm}. At lower magnetization, a stronger O-mode contribution can lead to mixed-mode emission and more variable polarization fractions \citep{Plotnikov_2019}.

Variability in the upstream magnetic field could, in principle, produce time-dependent PPA behavior. For example, current-driven kink instabilities can distort an initially toroidal magnetic field into a non-axisymmetric structure, causing the shock to encounter rapidly varying field orientations \citep{2016MNRAS.456.1739B}. Since the maser emission is governed by the local upstream field direction, such variations can lead to corresponding changes in the observed PPA. In contrast, secular changes in the outflow orientation, for instance due to precession of the central engine or surrounding disk, would be expected to produce smooth PPA evolution over longer timescales (days to months) \citep{Sridhar_PIC, 2020Natur.586..693L}.

Both variability scenarios described above can generate time-dependent PPA swings; they generally predict smooth evolution or variability tied to gradual changes in the upstream field geometry. They do not naturally reproduce the abrupt, burst-to-burst PPA changes observed in \ronefourseven on millisecond-to-second timescales (Figure~\ref{fig:ppa60382}) without invoking rapidly varying or finely tuned external conditions. Moreover, variations in the upstream field orientation alone do not necessarily predict correlated changes in the polarization fractions, making it difficult to simultaneously explain the observed PPA variability and episodes of enhanced circular polarization \ref{fig:60683_bottom}.

\subsubsection{Magnetospheric origin}

A second major class of FRB models places the emission region inside the magnetosphere of a strongly magnetized neutron star. In this scenario, coherent radiation is generated very near the open magnetic field lines emerging from the polar-cap region or from the magnetospheric current sheet outside the light cylinder region \citep{crab_putside_LC_2019}.

For non-aligned rotators, in which the angle between the magnetic and spin axes is large, RVM naturally produces the classic S-shaped PPA swing \citep{Beniamini_2025}, as recently observed in FRB~20221022A \citep{McKinven-swing_2024} and in several bursts of \rsixtyseven \citep{2024ApJ...972L..20N}. In addition, \citet{Wang__2022} argue that variations in the PPA across the burst envelope may require more complex magnetic-field configurations along LOS, beyond a simple dipolar geometry.

In contrast, for nearly-aligned rotators, the magnetic and spin axes are separated by only a small angle, so the emission beam remains continuously within the observer’s field of view (\citep{Beniamini_2025}; see their Fig. 1 and 2).  
This configuration provides a natural explanation for the highly stable PPAs observed in repeaters such as \rone \citep{Michilli_2018} and \rthree \citep{bethapudi2025constrainingoriginlongterm}, and small variations in \rsixtyseven \citep{Hilmarsson_2021}, where minimal geometric evolution is expected from burst to burst. However, the situation is markedly different for our source. We observe large, rapid, and apparently stochastic PPA jumps, spanning timescales from $\sim$ 15~ms to tens of minutes in the NRT data and up to $\sim$ 2.5~hr in the Effelsberg observations. Such behavior is difficult to reconcile with a single, fixed emission region in an aligned rotator.

These observations instead suggest (1) multiple active emission sites within the magnetosphere --- each sampling a different local magnetic-field orientation --- (2) an emission mechanism different from what occurs in radio pulsars, or (3) strong magnetospheric propagation effects, such as birefringent mode coupling or generalized Faraday rotation. For example, \citep{lyutikov_2025} propose the Anderson self-localization scenario in pair plasma, in which localized emission structures can acquire different polarization angles (see their Figure 16). On the other hand, birefringent mode coupling \citep{Oswald_2023_partial_coherence} or generalized Faraday rotation \citep{kumar_2023_GFR}  can imprint rapid PPA variability even when the emission geometry remains unchanged. Consequently, the standard aligned-rotator magnetospheric models may require significant revision to account for the degree of PPA variability exhibited by our source.

\subsubsection{Comparisons to Crab giant pulses and magnetar single pulses}

The complex polarization behavior of \ronefourseven resembles that of the giant pulses of the Crab pulsar and of some radio-emitting magnetars. High-time-resolution polarization studies of the Crab giant pulses reveal highly complex polarization behavior in both the PPA and the linear and circular polarization fractions \citep{Jessner_2010, Hankins_2016}. Figure~8 of \citep{Hankins_2016} shows individual emission entities of a pulse that displays nearly 100\% linear and some circular polarization, and the corresponding PPAs vary significantly, lacking any ordered RVM. Furthermore \citep{Jessner_2010} report that the interpulse PPAs are confined to $\pm\ 30^{\circ}$, while the main pulse PPAs span the $0^{\circ} - 180^{\circ}$ range. 
These are attributed to multiple short-lived, coherently emitting structures rather than to a single stable emission geometry. Such complex, short-lived coherent emission structures are consistent with coherent emission from reconnection-driven plasmoid activity in the current sheet beyond the light cylinder, where successive plasmoid mergers generate short-lived coherent radio bursts \citep{crab_putside_LC_2019}.

The average radio emission of magnetars is composed of numerous narrow, bright and highly variable spikes or subpulses \citep[e.g.][]{10.1111/j.1365-2966.2007.11622.x,2012MNRAS.422.2489L}. The PPA is nearly constant across an individual spike, while the PPA values of successive spikes follow the broader trend traced by the average profile \citep{10.1111/j.1365-2966.2007.11622.x,2012MNRAS.422.2489L}. In the RVM model, this behavior arises naturally because a narrow spike samples only a small interval of the underlying PPA variation with rotational phase. For \ronefourseven, however, the rotational period and phase of individual bursts are unknown, so the burst-averaged PPAs provide a phase-unwrapped sampling of this underlying variation. Thus, an average profile with a relatively small PPA swing would be expected to produce a narrow distribution of burst PPAs, whereas a larger swing would produce a broader distribution. This motivates using the PPA measurements themselves to search for a rotational signature; even if burst arrival times are stochastic, the PPA may retain information about the source's rotational phase and thus reveal an underlying periodicity \citep{kaustabh_ppa_2025}. However, we find no significant periodic modulation in the PPA time series within individual epochs.

Temporal variability of magnetar polarization can occur on timescales ranging from days to months and has been attributed to changes in the emission region, magnetospheric plasma and current configuration, or viewing geometry \citep[e.g.][]{Dai_et_al_J1810_2019, Champion_et_al_J1818_2020, Lower_et_al_J1818_2021,lower_xray_2023}. 
A particularly striking example is \xteeighteen, where high-cadence observations following the 2018 X-ray outburst revealed systematic evolution of the PPA swing, including reversals in its gradient in the interpulse (IP); \citet{Desvignes_2024} interpreted these variations as evidence for free precession, with the changing orientation of the star modifying the line-of-sight geometry relative to the magnetic axis. 
These observations show that magnetar PPA behavior can evolve substantially between epochs, from relatively narrow swings to broader and more complex distributions. 
 In \ronefourseven, the broader and narrower $\mathrm{PPA}_0$ distributions observed across different activity phases may likewise reflect longer-term changes in its magnetospheric or viewing geometry, although such evolution cannot account for the rapid burst-to-burst PPA variations.

Propagation effects in the magnetosphere can also modify the observed PPA. In \xteeighteen, the broader PPA distributions observed in some emission components were accompanied by stronger circular polarization \citep{10.1111/j.1365-2966.2007.11622.x}. 
\citet{Lower_2024} observed a period of linear-to-circular polarization conversion in \xteeighteen, coincident with a substantially broader and branching distribution of PPAs; the frequency- and pulse-phase-dependent behavior was interpreted in terms of birefringent propagation and coupling between polarization modes within or near the magnetosphere. 
Narrowband radio bursts from the magnetar 1E~1547.0$-$5408 at higher frequencies show a broad spread in polarization angles, which has also been attributed to emission from multiple active regions or propagation effects within the magnetosphere \citep{lower2026transientnarrowbandradiobursts}. The large burst-to-burst PPA variations in \ronefourseven may likewise point to multiple emission regions and/or magnetospheric propagation as possible contributors to its PPA diversity.

\section{Conclusion}
\label{section6} 
We present a detailed polarimetric study of 190 bursts from \ronefourseven, spread over $\sim$\ 1 year time period, focusing on the temporal behavior of the polarization position angle on timescales from milliseconds to hours. Our main findings are:
 
\begin{enumerate}

\item We find strong PPA variability, with burst-to-burst differences reaching up to $90^\circ$ and occurring on timescales ranging from $\sim10$–$15$\,ms to a few hours. The $\Delta\mathrm{PPA}$ distribution is symmetric about zero and shows no statistically significant dependence on burst separation timescales (e.g., $\leq 1$\,s versus $\geq 1$\,s). This indicates that the same underlying mechanism governs the PPA jumps in both closely spaced and widely separated bursts.

\item Despite this strong PPA variability, the emission remains predominantly highly linearly polarized. Across our full sample, $\sim81\%$ of bursts exhibit $L/I > 0.8$, while circular polarization is generally weaker, with $\sim16\%$ of bursts showing $|V/I| > 0.1$. These polarization fractions remain broadly consistent across three phases of enhanced burst activity, indicating stable overall polarization properties despite variations in burst rate. 

\item Large $\sim90^\circ$ jumps, as expected from orthogonal polarization modes, are rare. This suggests that the observed PPA variability arises from a broad, continuous distribution of polarization angles rather than from discrete mode switching.

\item Existing models reproduce some aspects of the observed behavior, but not all. External-shock scenarios generally predict stable PPAs or smooth temporal evolution tied to changes in the upstream magnetic-field geometry, and therefore do not naturally account for the observed rapid, burst-to-burst PPA variations. Magnetospheric models with emission from a single, fixed region in a dipolar field similarly struggle to reproduce the large-amplitude, stochastic PPA variations. The observations instead favor either multiple active emission regions within a complex magnetosphere or strong propagation effects, including birefringent mode coupling in the magnetosphere or plasma lensing in the surrounding source environment.

\end{enumerate}
Taken together, \ronefourseven combines a relatively stable Faraday environment with extreme, burst-to-burst PPA variability spanning milliseconds to hours. This combination highlights the polarization position angle as a powerful probe of the emission geometry and local plasma conditions in FRB sources. Continued high-time-resolution, broadband polarimetric observations, particularly during periods of high burst activity, will be essential for distinguishing between magnetospheric and external-shock scenarios and for constraining the physical origin of FRB emission.

\begin{acknowledgements}
We would like to thank the reviewer for the useful advice on improving the quality of the manuscript. We would also like to thank Maxim Lyutikov for pointing out some possible theoretical explanations of our results.
N.M. thanks Paz Beniamini for insightful discussions on the theoretical interpretation of our results and Mark Snelders for valuable discussions on polarimetry. L.G.S. is a Lise Meitner Group Leader and, together with N.M., acknowledges support from the Max Planck Society. This work was also carried out as part of the AstroFlash research group at McGill University, the University of Amsterdam, ASTRON, and JIVE, which is supported by a Canada Excellence Research Chair in Transient Astrophysics (CERC-2022-00009); an Advanced Grant from the European Research Council (ERC) under the European Union’s Horizon 2020 research and innovation programme (“EuroFlash”, Grant agreement No. 101098079); an NWO Vici grant (“AstroFlash”, VI.C.192.045); an NSERC Discovery Grant (RGPIN-2025-06681); an ERC Starting Grant (“EnviroFlash”, Grant agreement No. 101223057); and an NWO Veni grant (VI.Veni.222.295).

This work is based on observations obtained with the Nançay Radio Observatory, operated by the Paris Observatory and associated with the French Centre National de la Recherche Scientifique (CNRS). This work also uses observations obtained with the 100-m telescope of the Max-Planck-Institut für Radioastronomie (MPIfR) at Effelsberg. The Effelsberg EDD backend system is developed and maintained by the electronics division at MPIfR.

\end{acknowledgements}

\bibliography{reference}
\begin{appendix}
\section{Polarization Calibration}
\label{pol_cal_appendix}
In a linear-feed receiver, the Stokes parameters can be derived from the
auto- and cross-correlation:  
\begin{equation}
\begin{aligned}
I &= AA + BB \\
Q &= AA - BB \\
U &= 2\,CR \\
V &= -2\,CI \\  
\end{aligned}
\label{eq:Stokes_eq}
\end{equation}
where \(AA\) and \(BB\) are the auto-correlations of the two polarization
channels, and \(CR\) and \(CI\) are the real and imaginary parts of the
cross-correlation product, respectively.
Any instrumental delay between the polarization channels introduces
a phase shift in the cross term \(CR\) and \(CI\), which leads to distortions in Stokes
\(U\) and \(V\). 
To correct this, we used 0.2-second scans of a 3.33-Hz pulsed noise diode,
recorded during each observing session. These calibration scans were
performed in all 8 sub-bands and later stitched into a single solution.  

The phase angle of the noise diode as a function of frequency is then calculated as  

\begin{equation}
   \mathrm{PA}\ (\nu) =\frac{1}{2}\ \arctan \left( \frac{CR}{CI} \right) 
\end{equation}

and a linear slope \(\mathcal{S}\) is fitted to this curve.  
The fitted slope, with units of rad Hz\(^{-1}\), is converted into a delay \(\mathcal{\tau}\) using  

\[
\mathcal{\tau} = \frac{\mathcal{S}}{\pi}.
\]

Typical delays between the polarization channels vary from 0.3 to 0.5~ns from one observing day to another. 
To correct for the instrumental delay between polarization channels, a phase correction is applied to the cross-coherence products, $CR$ and $CI$. The corrected cross products are computed as:

\begin{equation}
CR_{\rm corr} = Re \big(D\big)
\end{equation}

\begin{equation}
CI_{\rm corr} = Im \big(D\big)
\end{equation}

\begin{equation}
where\
D = \Big[ \big(CR + i\, CI\big) \, e^{-2 \pi i \nu \tau} \Big]
\end{equation}
This procedure removes the relative delay between the polarization hands.  

After the initial delay correction, small residual phase offsets were observed across channels in the noise diode signal. These residuals can be further corrected by introducing an additional channel-dependent phase term, $\phi_{\rm res}$:

\begin{equation}
CR_{\rm corr}= Re \big(D_{\rm phase}\big)
\end{equation}

\begin{equation}
CI_{\rm corr} = Im \big(D_{\rm phase})
\end{equation}
\begin{equation}
    where\ D_{\rm phase} = \Big[ \big(CR + i\, CI\big) \, e^{-2 i (\pi \nu \tau + \phi_{\rm res})} \Big]
\end{equation}

Finally, once the auto-correlations and instrumental-corrected cross products are obtained following the above procedure, the full Stokes parameters can be reconstructed using Equations~\ref{eq:Stokes_eq}. Thereafter, we also removed the parallactic angle effect from the bursts at each epoch.

\section{Comparison of PPA Measurements Using $RM_{\rm epoch}$ and $RM_{\rm burst}$}
\label{ppa_rm_comparison}

To assess whether the choice of RM used for Faraday correction affects the measured PPAs, we compare the burst-averaged PPAs obtained using the epoch-averaged rotation measure, $RM_{\rm epoch}$, with those obtained using the burst-specific rotation measure, $RM_{\rm burst}$. For each burst, the calibrated Stokes parameters were independently corrected for
Faraday rotation using the two RM measurements, and the corresponding
burst-averaged PPA was calculated using the same procedure described
in Section~3.3.

\begin{figure}
    \centering
    \includegraphics[width=\columnwidth]{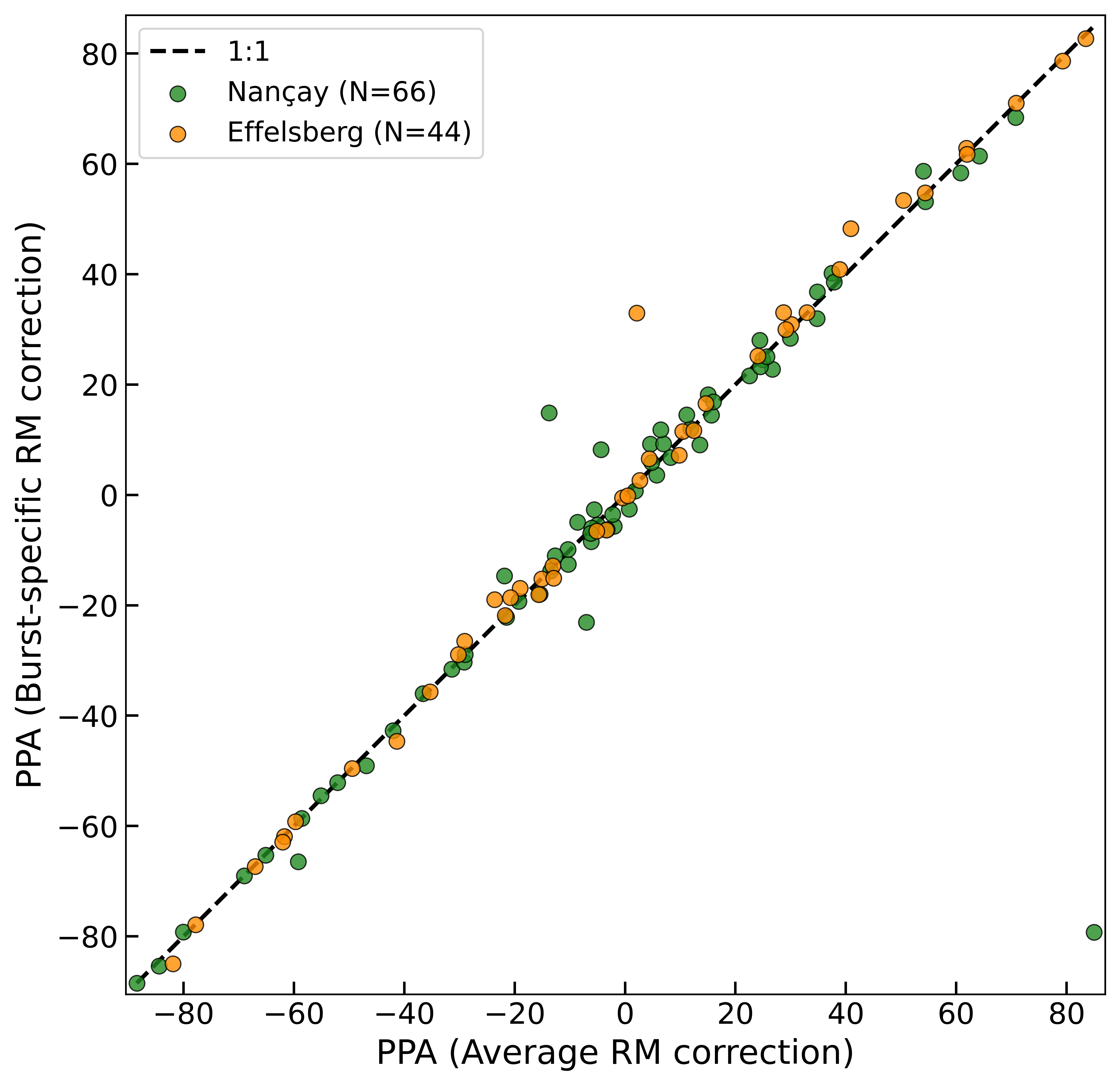}
    \caption{Comparison of burst-averaged polarization position angles obtained after Faraday correction using $RM_{\rm epoch}$, and $RM_{\rm burst}$. The dashed line represents the one-to-one relation. Most measurements show close agreement, with a small number of outliers discussed in Appendix \ref{ppa_rm_comparison}. The sample includes Nançay observations from MJDs 60382, 60515, and 60683, and Effelsberg observations from MJD 60384.}
    \label{fig:ppa_rm_comparison}
\end{figure}

Figure~\ref{fig:ppa_rm_comparison} shows the resulting comparison, with the dashed line representing the one-to-one relation. Most measurements lie close to this relation, indicating that the choice of RM correction has little effect on the measured PPA for most bursts. The apparent extreme outlier
arises from the $180^\circ$ ambiguity in PPA, while four additional outliers have burst-specific RMs that differ appreciably from their respective epoch averages, with relatively large RM uncertainties in several cases. Overall, the agreement indicates that the burst-to-burst PPA variability reported in this work is not a consequence of using the epoch-averaged RM for Faraday correction.

\begin{table*}
\caption{Polarization properties from repeaters with published PPA statistics}
\label{tab:frbtable}
\renewcommand{\arraystretch}{1.2}
\small

\begin{threeparttable}

\begin{tabularx}{\textwidth}{l c X X X X X X X}
\hline \hline
Source & $\mathrm{RM}_{\mathrm{rest}}$ & $\Delta$RM & $\Delta$PPA & $\Delta$PPA & DM$_{\rm Host}$ & PRS & Reference\\ 
& (rad~m$^{-2}$) & (rad~m$^{-2}$) & within bursts & between bursts & (pc~cm$^{-3}$) & (erg~s$^{-1}$~Hz$^{-1}$) & \\ 
\hline

FRB~20121102A  
& $4.4\times10^{4}$ to $1.5\times10^{5}$
& $10^{5}$ 
& flat 
& flat 
& 55--225 
& $\sim 2\times10^{29}$ 
& \tnote{1,2,3} \\

FRB~20201124A  
& $\sim -1073$ to $-434$
& $\sim 500$ 
& flat\tnote{a}
& $20^\circ$--$30^\circ$ 
& 150--220 
& $1.2\times10^{29}$ 
& \tnote{4,5,6,7,8} \\

FRB~20180916B 
& $\sim -128$ to $-53$
& $\sim 70$ 
& flat 
& flat 
& $<70$ 
& -- 
& \tnote{6,9,10} \\

\ronefourseven 
& $\sim 434$ to $472$
& $\sim 20$ 
& $\sim$flat~or~slanted
& $\sim$small to $\gtrsim 90^\circ$
& 150 
& $2.2 \times 10^{28}$ 
& \tnote{11,12} \\

FRB~20190417A  
& $5038$ to $6441$
& $\sim 1000$ 
& -- 
& -- 
& $>1228.7$ 
& $7.4 \times 10^{28}$ 
& \tnote{13,14} \\

\hline
\end{tabularx}

\begin{tablenotes}

\footnotesize
\item[a]{also varying (RVM), orthogonal}
\item[] 
[1] \cite{2023ATel15980....1F}, 
[2] \cite{Michilli_2018}, 
[3] \cite{Chatterjee_2017, chen2022comprehensiveobservationalstudyr1, 2017ApJ...834L...7T},[4] \cite{Xu_2022}, 
[5] \cite{liu2025polarizationpositionangleswing}, 
[6] \cite{10.1093/mnras/stab2936}, 
[7] \cite{2024ApJ...972L..20N}, 
[8] \cite{Ravi_2022}, [9] \cite{Bethapudi_2025}, 
[10] \cite{Marcote_2020},[11] \cite{chen2025hostgalaxyhyperactiverepeating}, 
[12] \cite{Bruni_2025},[13] \cite{moroianu2025milliarcsecondlocalizationassociatesfrb}, 
[14] \cite{Feng_2025}.
\end{tablenotes}

\end{threeparttable}
\end{table*}

\begin{figure*}[h!]
    \centering

    \begin{subfigure}{0.48\textwidth}
        \centering
        \includegraphics[width=\linewidth]{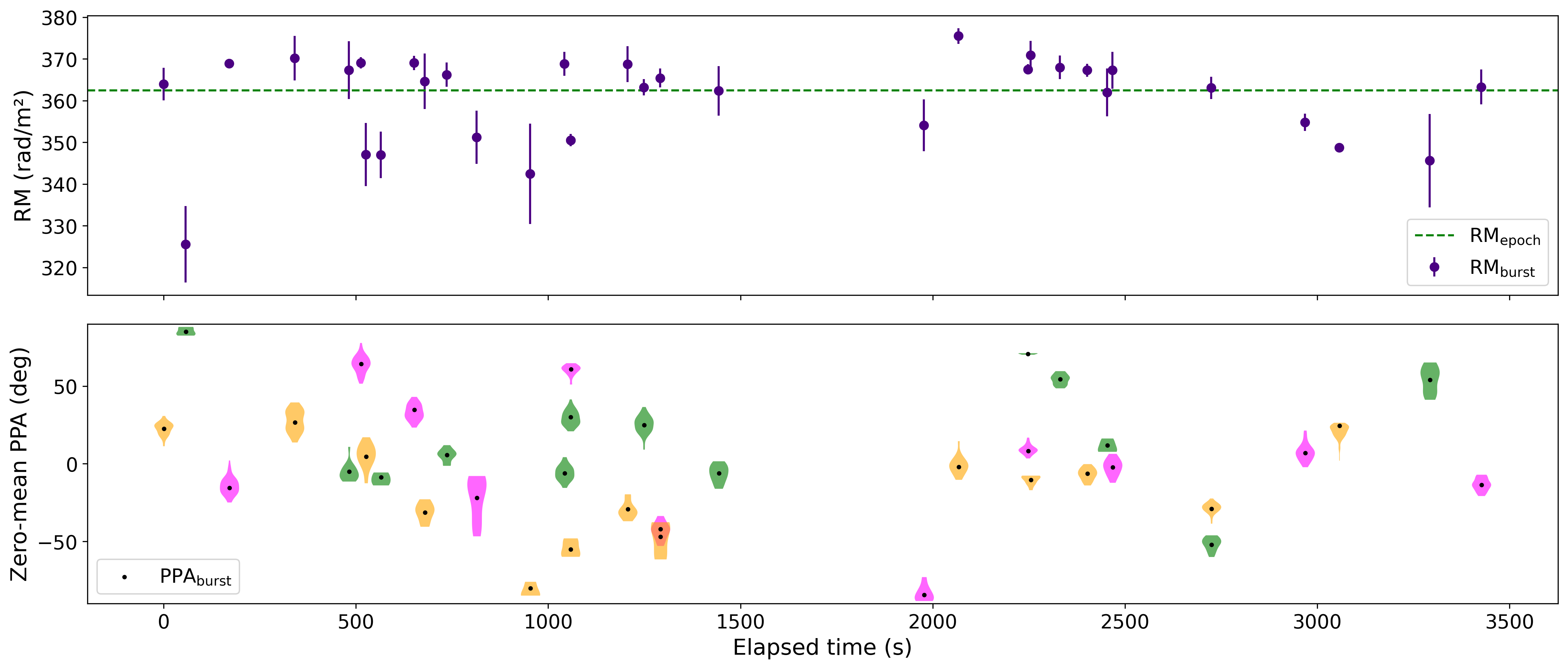}
        \label{fig:rm60382}
        \vspace{0.3cm}
        
        \includegraphics[width=\linewidth]{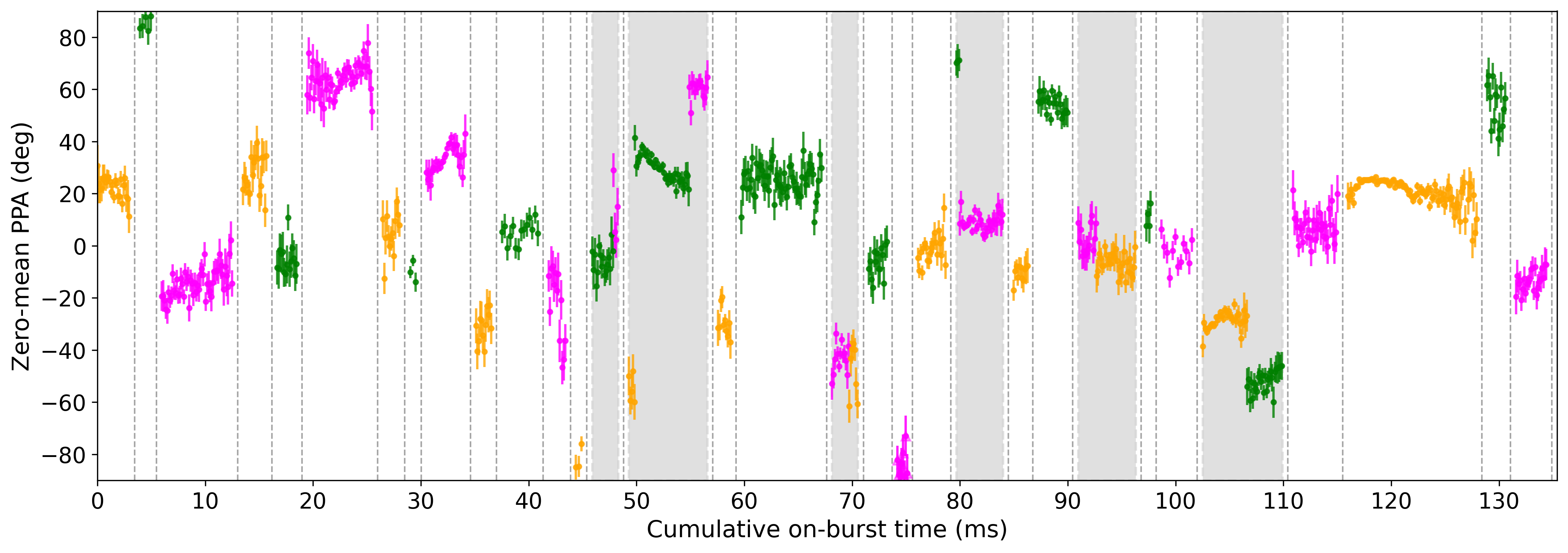}
        \caption{MJD~60382 (P1)}
        \label{fig:ppa60382}
    \end{subfigure}
    \hfill
    \begin{subfigure}{0.48\textwidth}
        \centering
        \includegraphics[width=\linewidth]{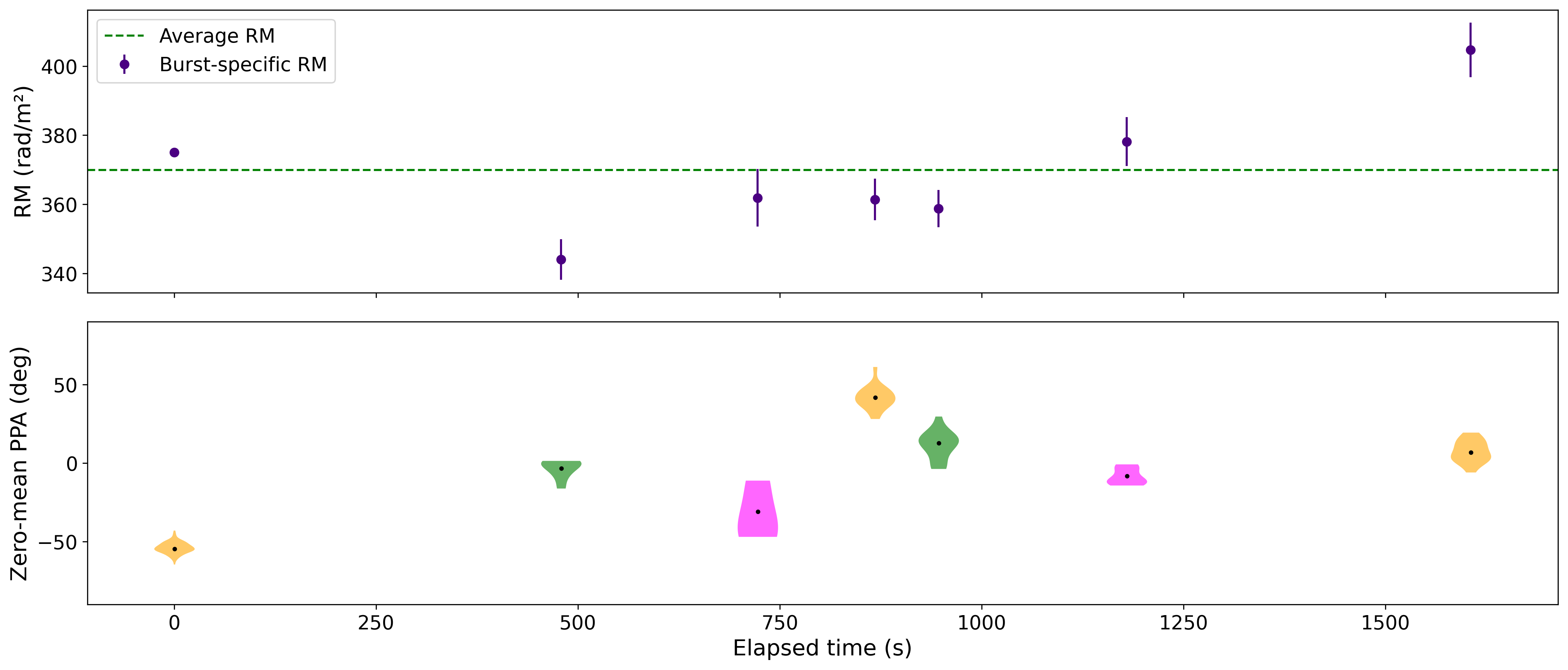}
        
        \vspace{0.3cm}
        
        \includegraphics[width=\linewidth]{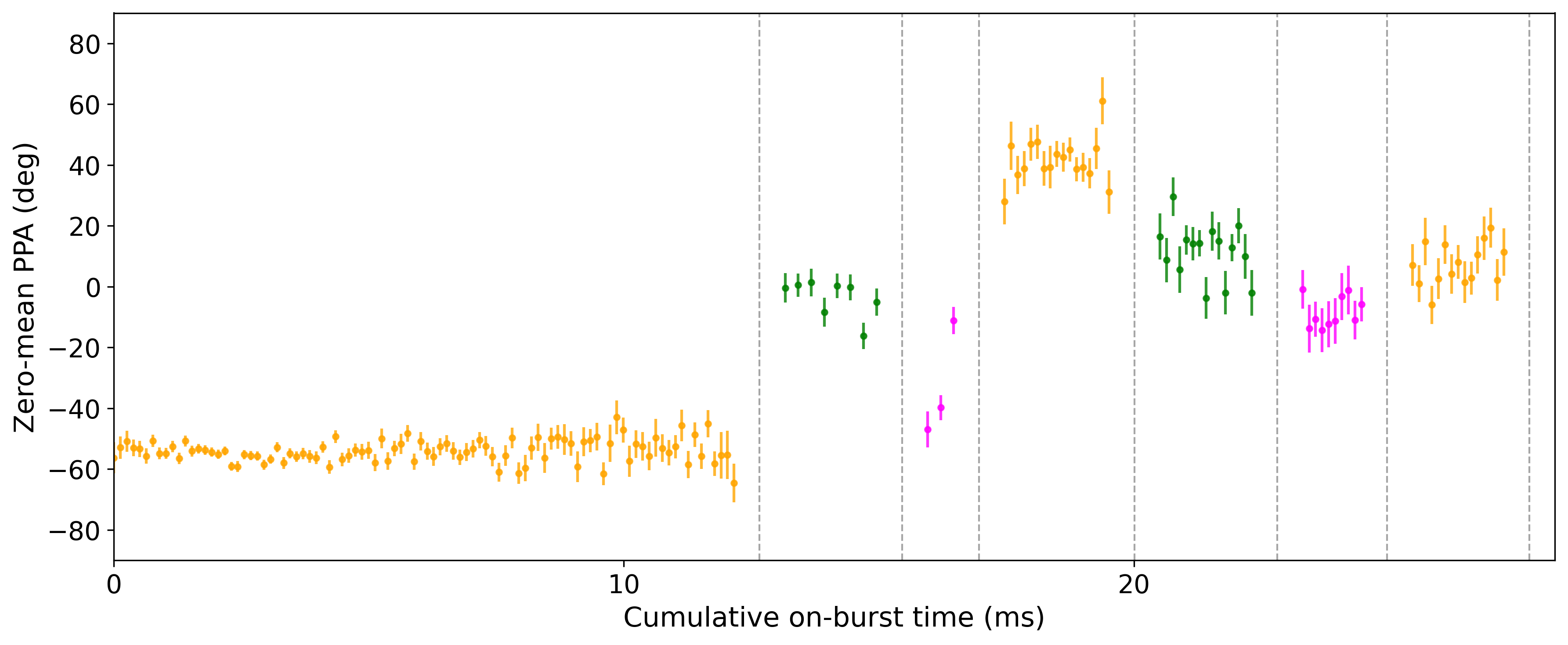}
        \caption{MJD~60385 (P1)}
        \label{fig:ppa60385}
    \end{subfigure}
    \hfill

    \caption{MJD~60382 and MJD~60385. Each panel shows RM–PPA behavior (top) and polarization fractions (bottom).}
    \label{fig:ppa_compare_P1_1}
\end{figure*}

\begin{figure*}[h!]
    \centering
    \begin{subfigure}{0.48\textwidth}
        \centering
        \includegraphics[width=\linewidth]{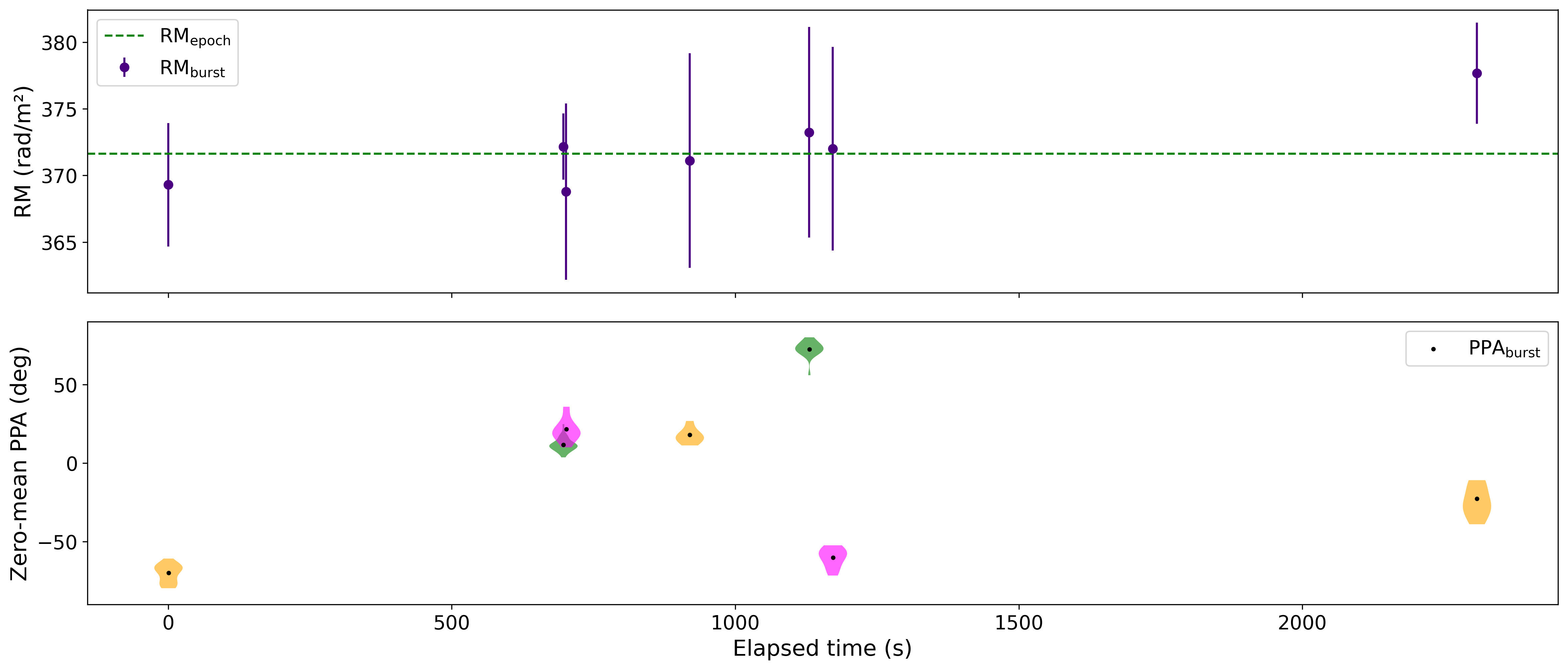}
        
        \vspace{0.3cm}
        
        \includegraphics[width=\linewidth]{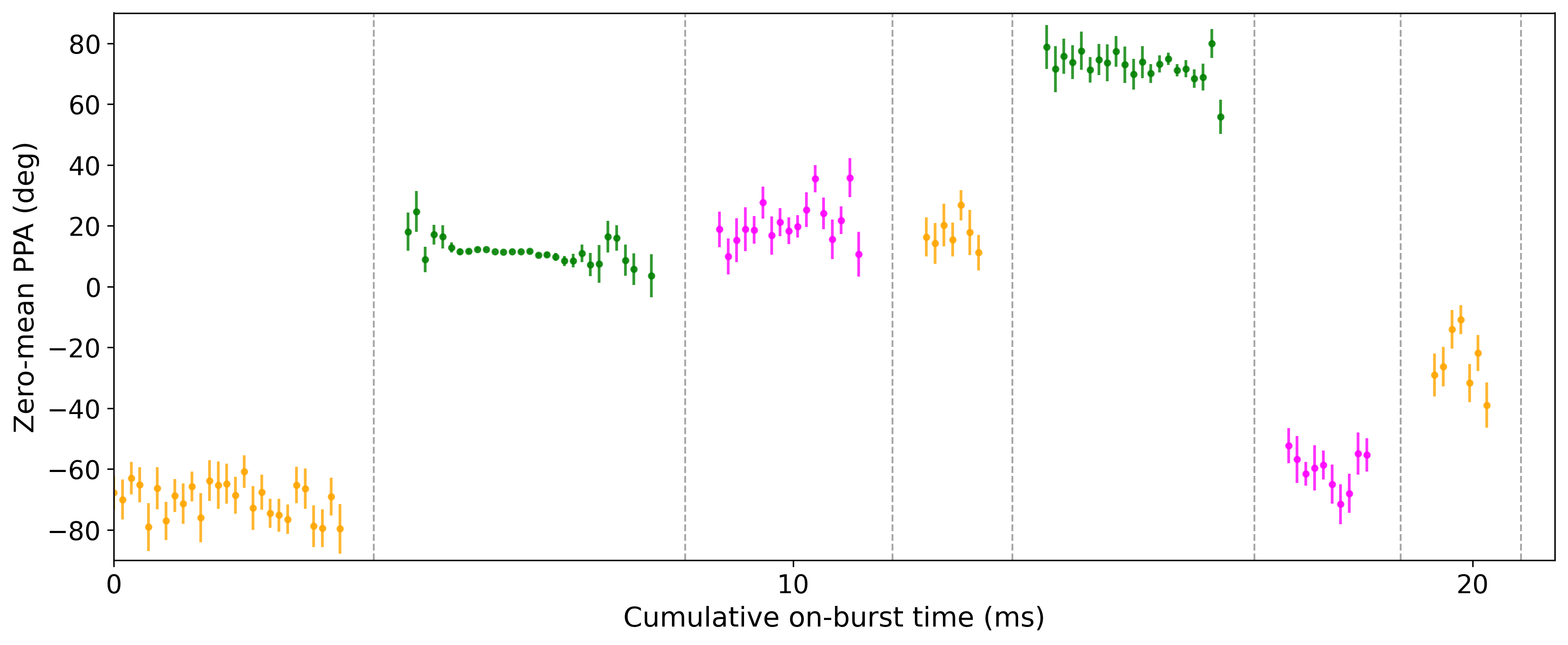}
        \caption{MJD~60386 (P1)}
        \label{fig:ppa60386}
    \end{subfigure} 
    \hfill
    \begin{subfigure}{0.48\textwidth}
        \centering
        \includegraphics[width=\linewidth]{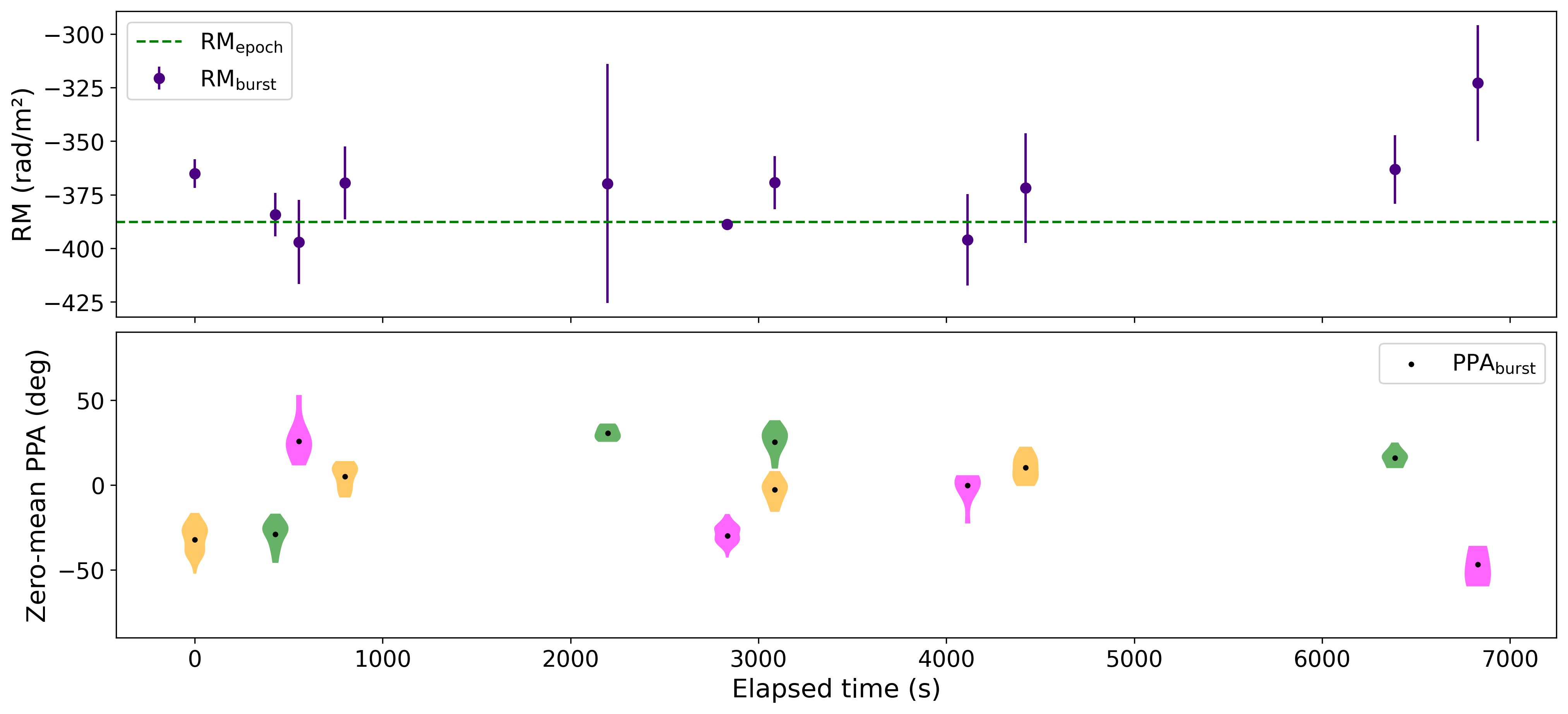}
        \vspace{0.3cm}
        
        \includegraphics[width=\linewidth]{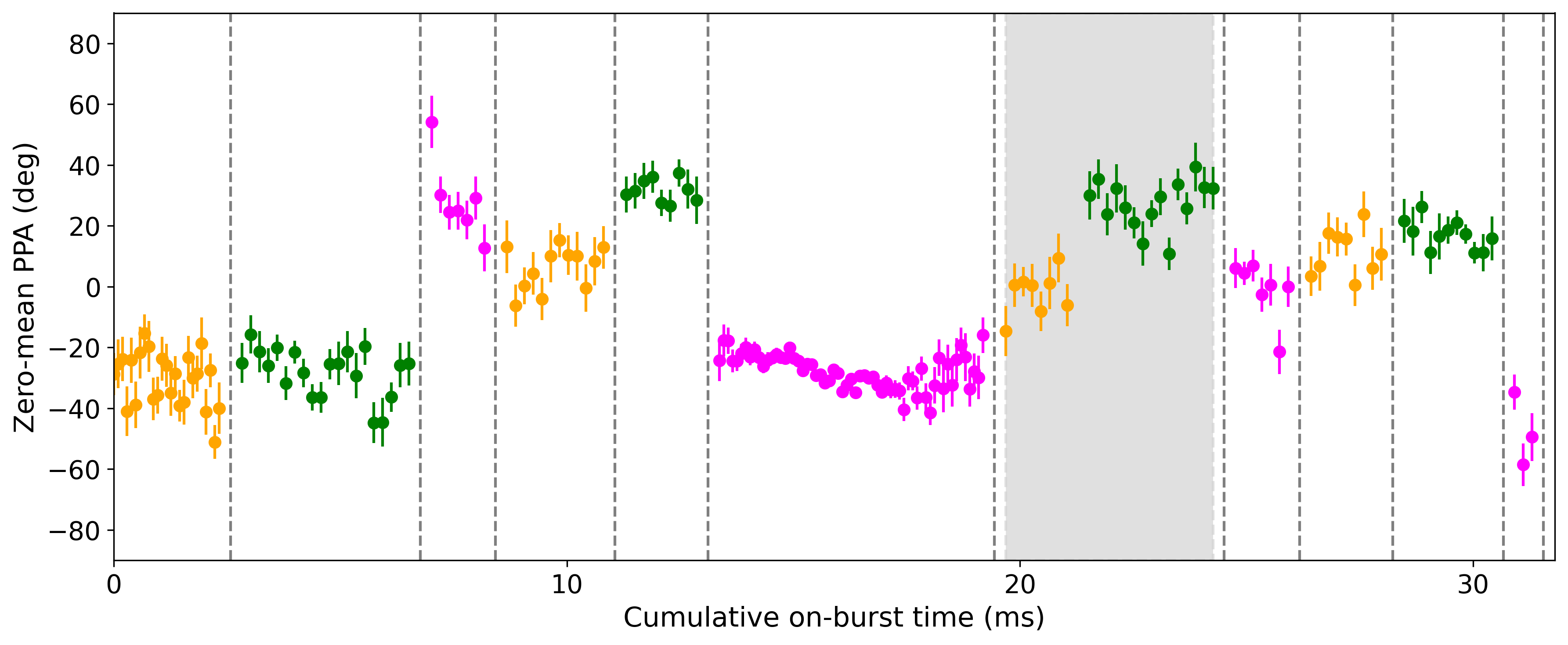}
        \caption{MJD~60398 (P1) EFF}
    \end{subfigure}
    
    \caption{MJD~60386 and MJD~60398. Each panel shows RM–PPA behavior (top) and polarization fractions (bottom).}
    \label{fig:ppa_compare_P1_2}
\end{figure*}

\begin{figure*}[h!]
    \centering

    \begin{subfigure}{0.48\textwidth}
        \centering
        \includegraphics[width=\linewidth]{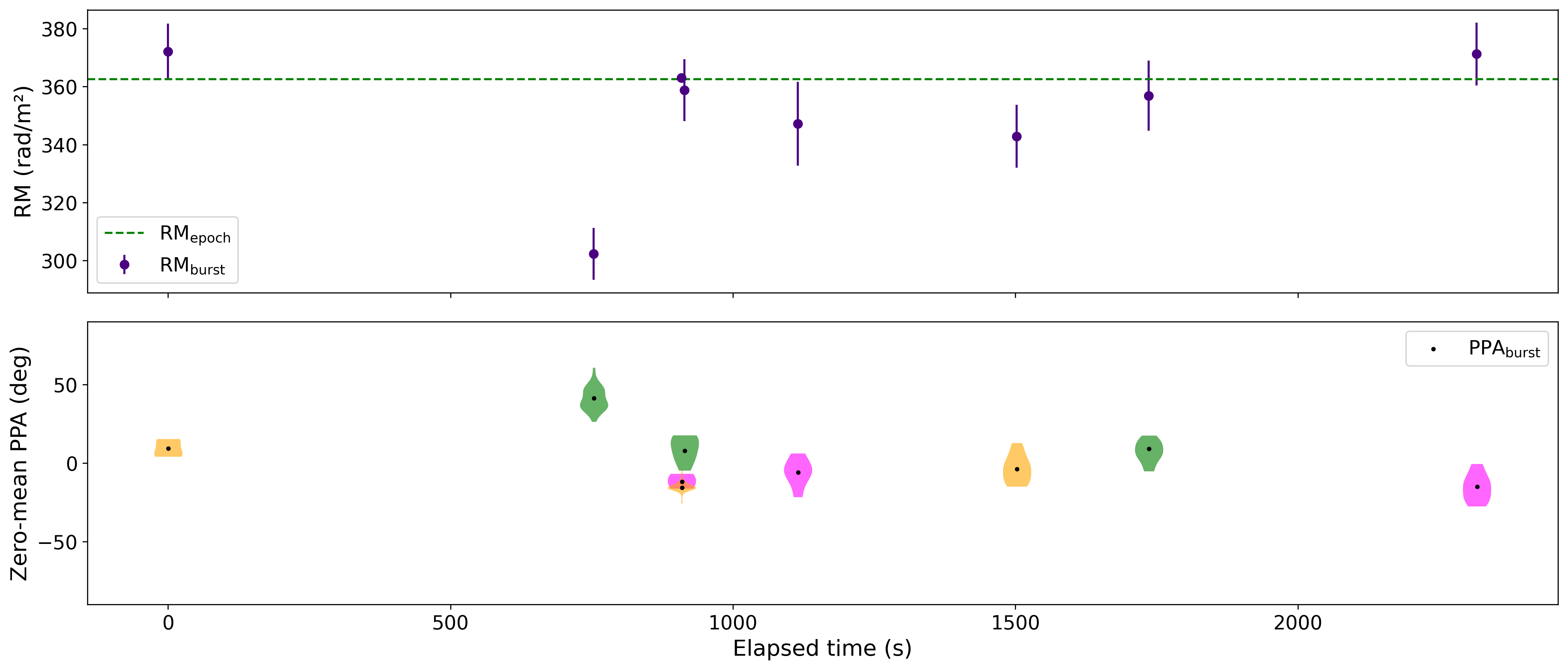}
        
        \vspace{0.3cm}
        
        \includegraphics[width=\linewidth]{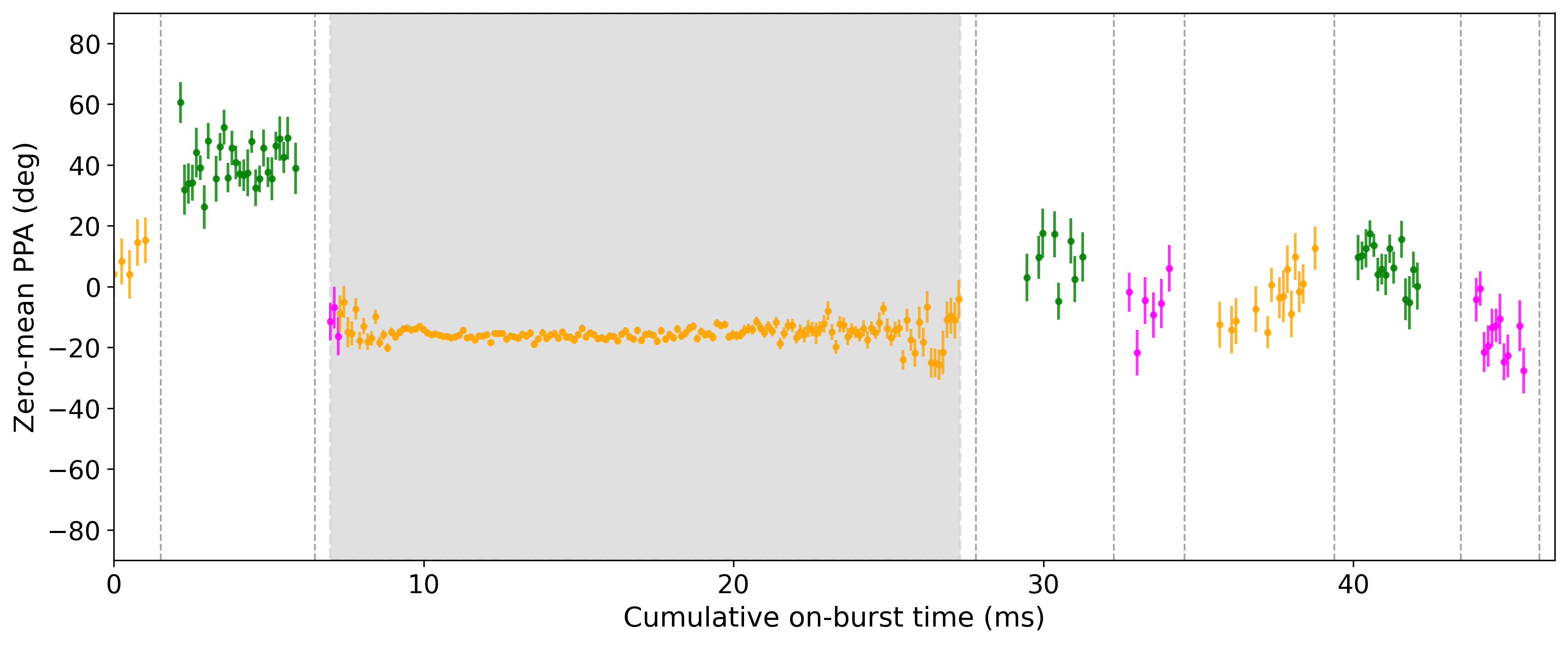}
        \caption{MJD~60506 (P2)}
        \label{fig:ppa60506}
    \end{subfigure}
    \hfill
    \begin{subfigure}{0.48\textwidth}
        \centering
        \includegraphics[width=\linewidth]{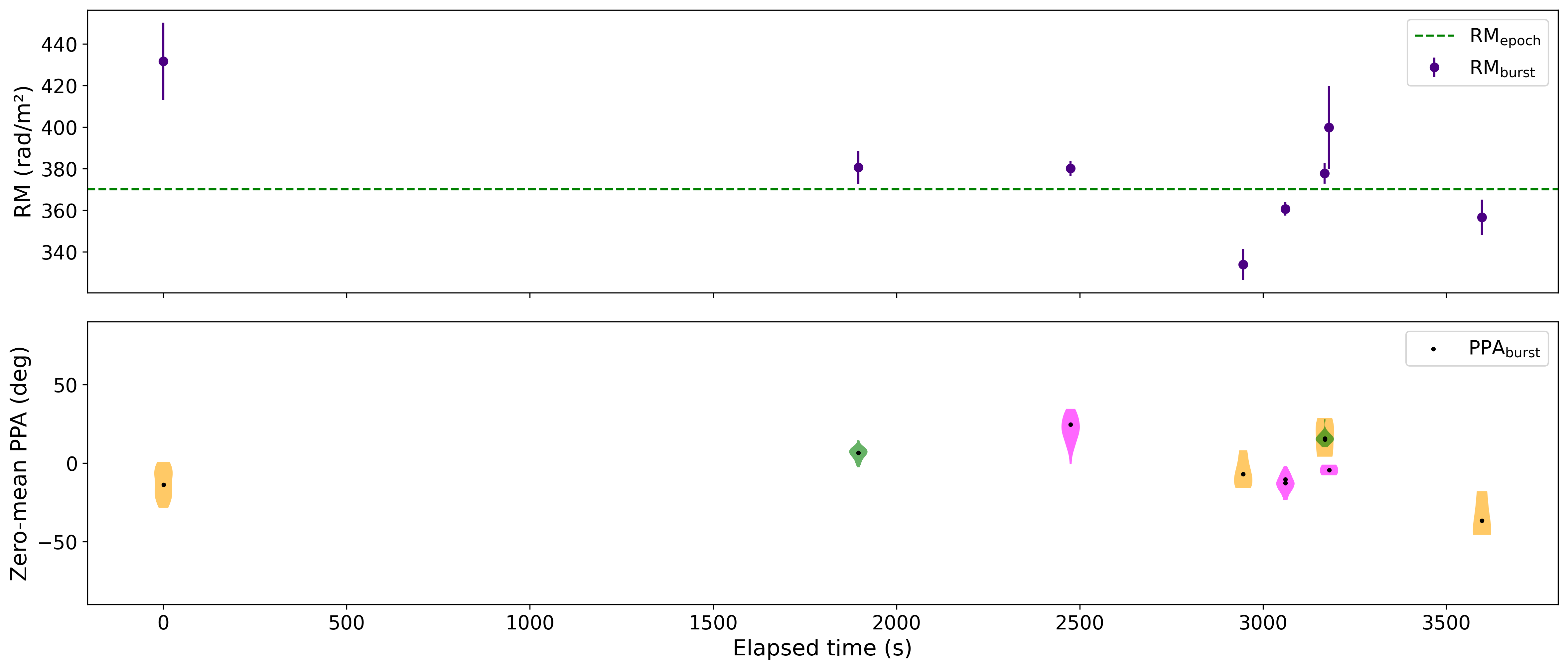}
        
        \vspace{0.3cm}
        
        \includegraphics[width=\linewidth]{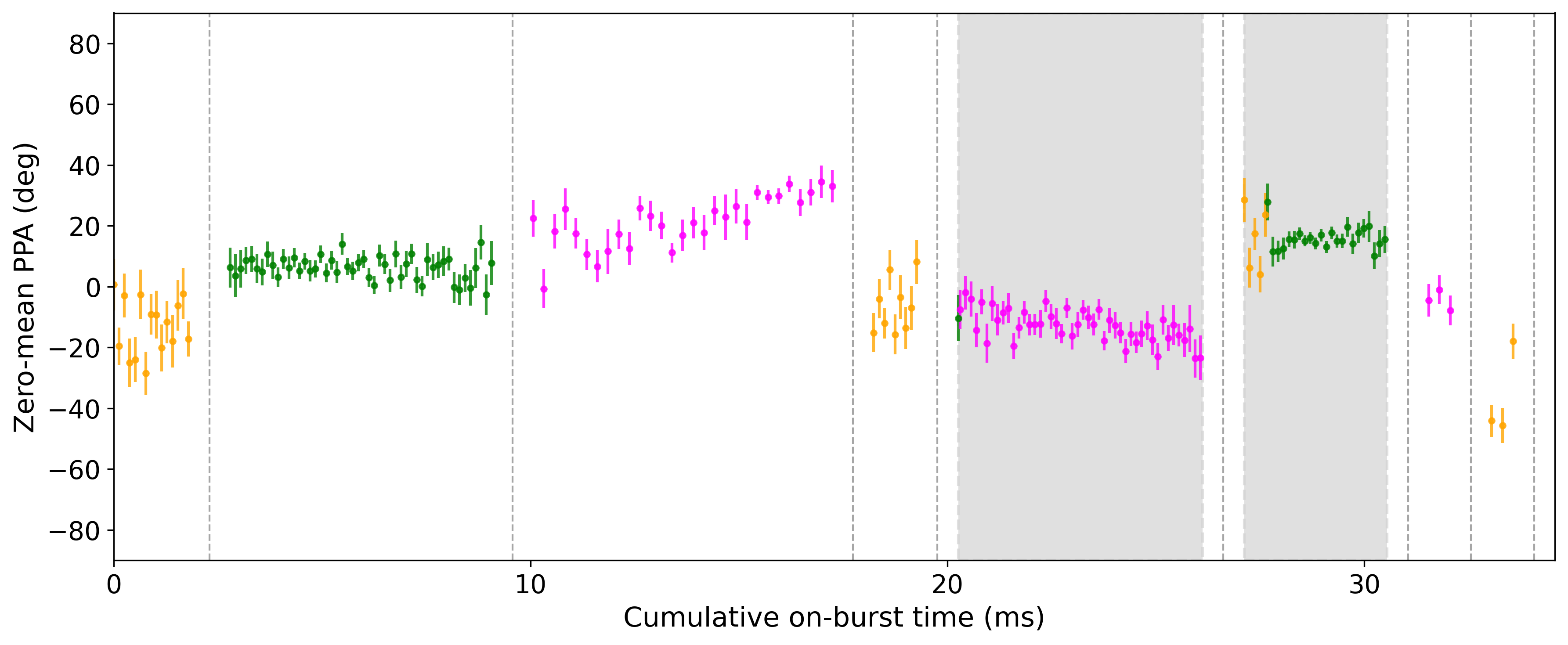}
        \caption{MJD~60515 (P2)}
        \label{fig:ppa60515}
    \end{subfigure}

    \caption{MJD~60506 and MJD~60515. Each panel shows RM–PPA behavior (top) and polarization fractions (bottom).}
    \label{fig:ppa_compare_P2}
\end{figure*}

\begin{figure*}[h!]
    \centering

    \begin{subfigure}{0.48\textwidth}
        \centering
        \includegraphics[width=\linewidth]{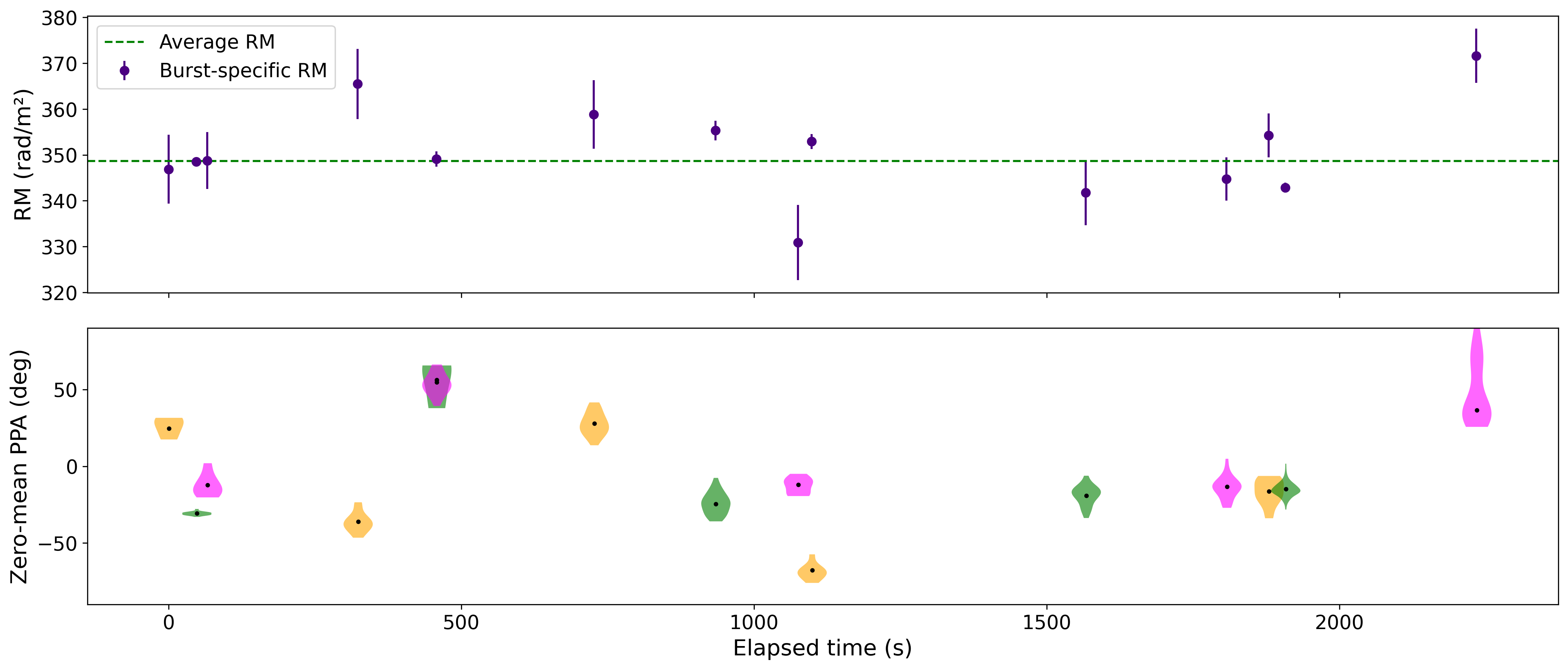}
        
        \vspace{0.3cm}
        
        \includegraphics[width=\linewidth]{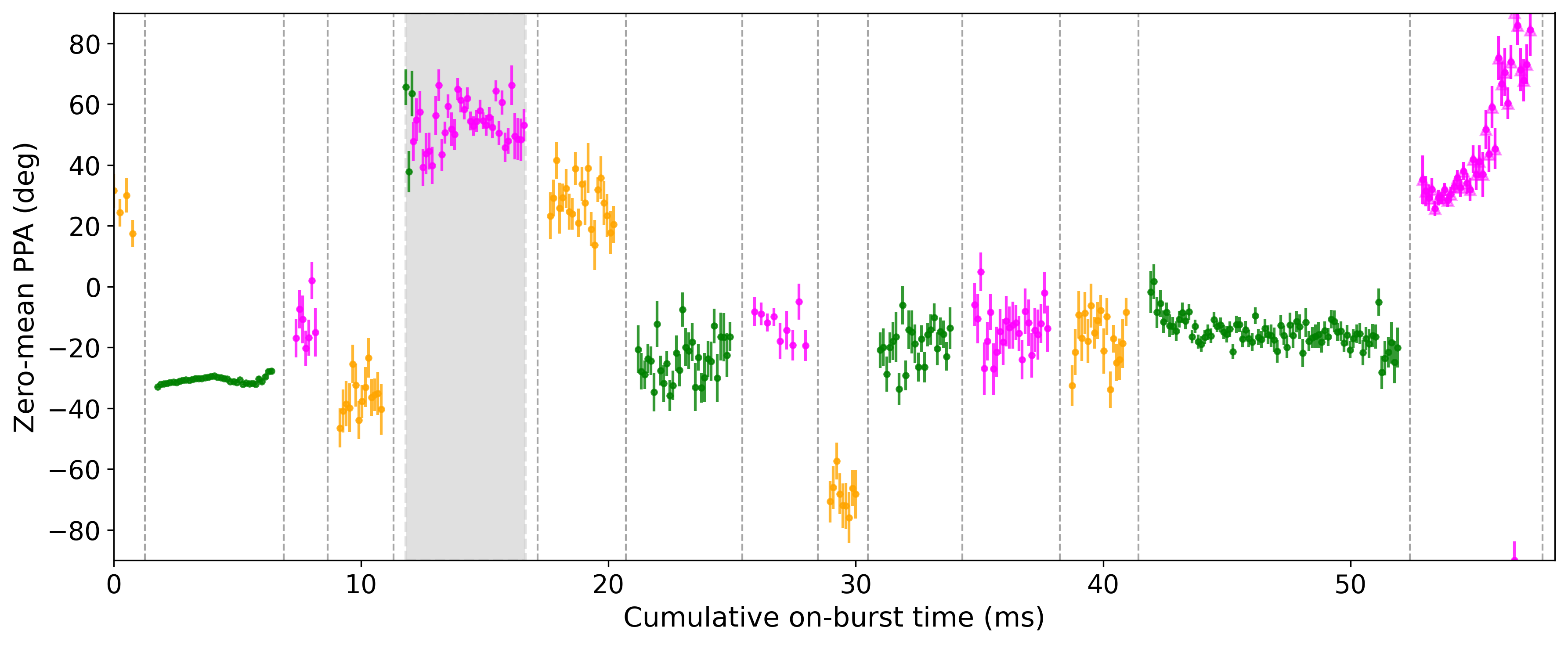}
        \caption{MJD~60673 (P3)}
        \label{fig:ppa60673}
    \end{subfigure}
    \hfill
    \begin{subfigure}{0.48\textwidth}
        \centering
        \includegraphics[width=\linewidth]{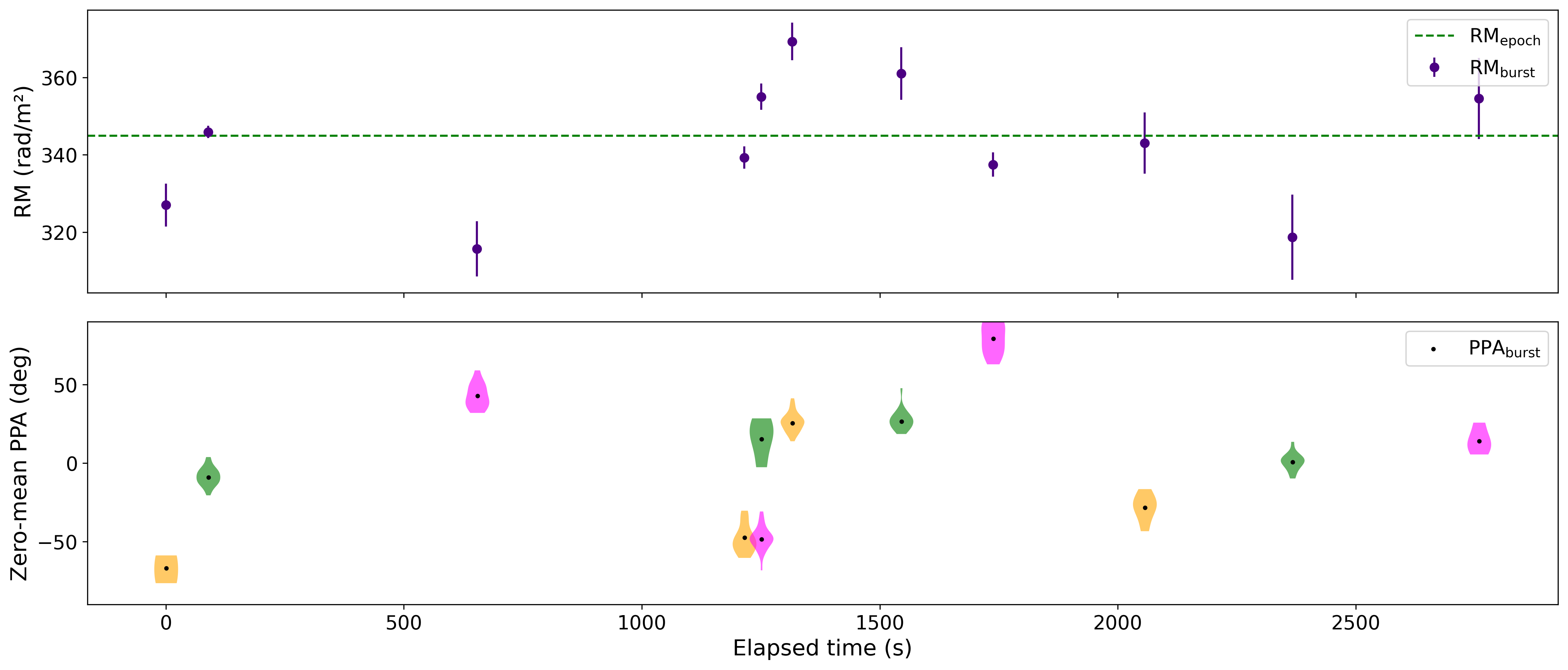}
        
        \vspace{0.3cm}
        
        \includegraphics[width=\linewidth]{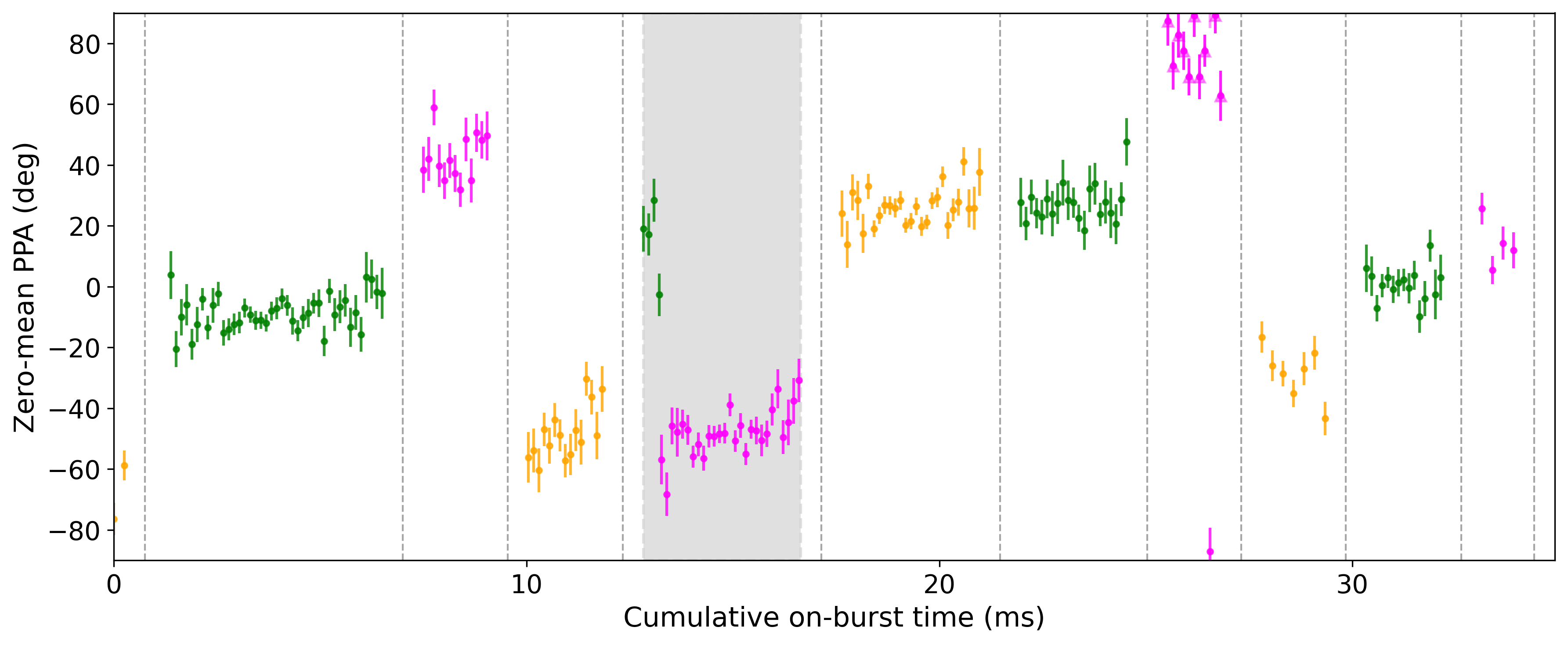}
        \caption{MJD~60687 (P3)}
        \label{fig:ppa60687}
    \end{subfigure}

    \caption{MJD~60673 and MJD~60683. Each panel shows RM–PPA behavior (top) and polarization fractions (bottom).}
    \label{fig:ppa_compare_P3}
\end{figure*}

\begin{figure*}[h!]
    \centering
    \includegraphics[width=0.9\linewidth]{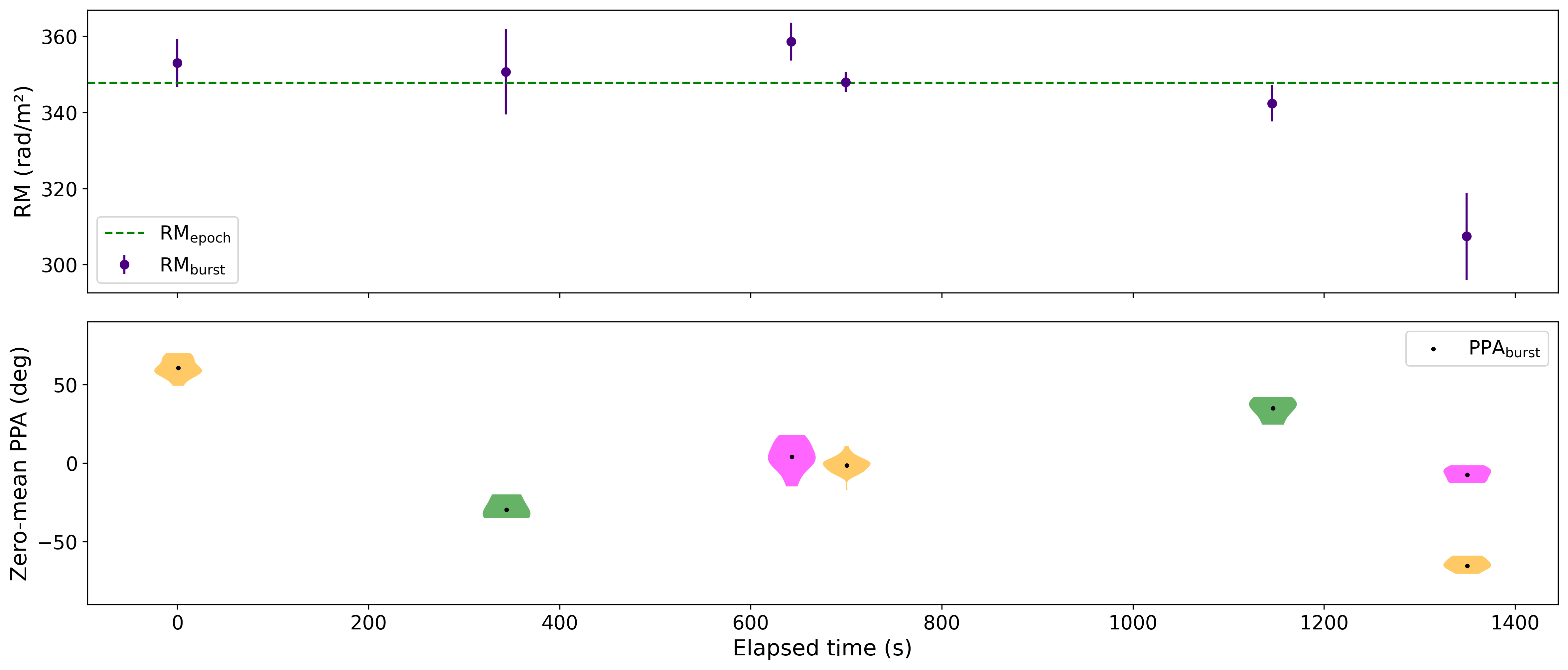}
    \label{fig:rm60692}
    \includegraphics[width=0.9\linewidth]{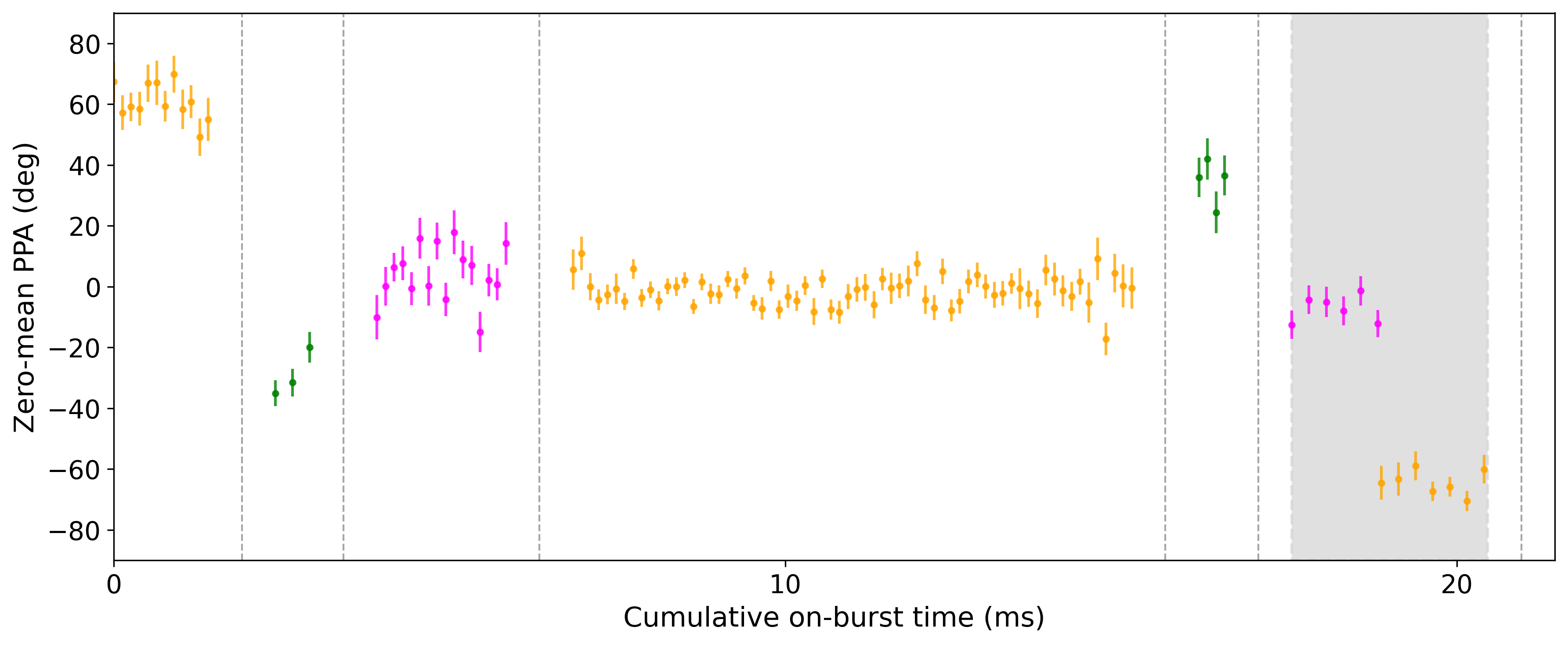}
    \caption{MJD~60692 (P3)}
    \label{fig:ppa60692}
\end{figure*}

\end{appendix}

\bibliographystyle{aa}

\end{document}